\documentclass[fleqn,usenatbib]{mnras}
\usepackage[T1]{fontenc}
\usepackage{ae,aecompl}

\usepackage{graphicx}	% Including figure files
\usepackage{amsmath}	% Advanced maths commands
\usepackage{amssymb}	% Extra maths symbols

\usepackage[dvipsnames]{xcolor}

\usepackage{amsmath}
\usepackage{float}
\usepackage{tikz}
\usepackage{listings}
\usepackage{caption}
\usepackage{animate}
\usepackage{subcaption}
\usetikzlibrary{positioning}
\usepackage[utf8]{inputenc}
\usepackage{setspace}
\usepackage{bm}
\usepackage{caption}
\usepackage{soul}
\usepackage[normalem]{ulem} %sout command

\newcommand*    \msun{{\,\mathrm{M}_{\odot}}}

\newcommand*    \pc{{\, \mathrm{pc}}}

\newcommand*    \mbh{M_\bullet}

\newcommand*    \ah{a_{\mathrm{h}}}
\newcommand*    \vrad{V_{\rm r}}
\newcommand*    \vtan{V_{\rm t}}

\title[Bound Eccentricity of black hole binaries]{
Eccentricity evolution of massive black hole binaries from formation to coalescence}

\author[A. Gualandris et al.]{%
Alessia Gualandris,$^{1}$\thanks{E-mail: a.gualandris@surrey.ac.uk}
Fazeel Mahmood Khan,$^{2,3}$
Elisa Bortolas,$^{4,5}$
Matteo Bonetti,$^{4,5}$
\newauthor
Alberto Sesana,$^{4}$
Peter Berczik$^{6,7}$
and Kelly Holley-Bockelmann$^{8,9}$
\\
$^{1}$Department of Physics, Faculty of Engineering and Physical Sciences, University of Surrey, Guildford GU2 7XH, UK \\
$^{2}$Department of Space Science, Institute of Space Technology, Islamabad 44000, Pakistan \\
$^{3}$Space and Astrophysics Research Lab (SARL), National Centre of GIS and Space Applications (NCGSA), Islamabad 44000, Pakistan\\ 
$^{4}$Dipartimento di Fisica “G. Occhialini”, Universit\`a degli Studi di Milano-Bicocca, Piazza della Scienza 3, IT-20126 Milano, Italy \\
$^{5}$INFN, Sezione di Milano-Bicocca, Piazza della Scienza 3, I-20126 Milano, Italy\\
$^{6}$National Astronomical Observatories and Key Laboratory of Computational Astrophysics, Chinese Academy of Sciences, \\ 
20A Datun Rd., Chaoyang District, 100012, Beijing, China \\
$^{7}$Main Astronomical Observatory, National Academy of Sciences of Ukraine, 27 Akademika Zabolotnoho St., 03680, Kyiv, Ukraine \\
$^{8}$Department of Physics and Astronomy, Vanderbilt University, Nashville, TN 37240, USA \\
$^{9}$Department of Physics, Fisk University, Nashville, TN 37208, USA
}

\date{}

\pubyear{2021}

\begin{document}
\label{firstpage}
\pagerange{\pageref{firstpage}--\pageref{lastpage}}
\maketitle

\begin{abstract}
Coalescing supermassive black hole binaries (BHBs) are expected to be the loudest sources of gravitational waves (GWs) in the Universe.  Detection rates for ground or space-based detectors based on cosmological simulations and semi-analytic models are highly uncertain. A major difficulty stems from the necessity to model the BHB from the scale of the merger to that of inspiral. Of particular relevance to the GW merger timescale is the binary eccentricity. Here we present a self-consistent numerical study of the eccentricity of BHBs formed in massive gas-free mergers from the early stages of the merger to the hardening phase, followed by a semi-analytical model down to coalescence.  We find that the early eccentricity of the unbound black hole pair is largely determined by the initial orbit. It systematically decreases during the dynamical friction phase. The eccentricity at binary formation is affected by stochasticity and noise owing to encounters with stars, but preserves a strong correlation with the initial orbital eccentricity. Binding of the black holes is a phase characterised by strong perturbations, and we present a quantitative definition of the time of binary formation. During hardening the eccentricity increases in minor mergers, unless the binary is approximately circular, but remains largely unchanged in major mergers, in agreement with predictions from semi-analytical models based on isotropic scattering experiments. Coalescence times due to hardening and GW emission in gas-poor non-rotating ellipticals are $\lesssim 0.5$ Gyr for the large initial eccentricities ($ 0.5 \leq e \leq 0.9$) typical of galaxy mergers in cosmological simulations. 
\end{abstract}


%%%%%%%%%%%%%%%%% BODY OF PAPER %%%%%%%%%%%%%%%%%%
\section{Introduction}
\label{sec:introduction}
Supermassive black holes (MBHs) with masses in the range $10^6-10^{10}\msun$ are commonly found at the centre of galaxies \citep[e.g.][]{FF2005, Kormendy2013} and are expected to be ubiquitous. In the hierarchical cosmological model of structure formation, MBHs come together during galactic mergers forming a bound pair \citep{Begelman1980}. Binaries of massive black holes (BHBs) may form in large numbers over cosmic time due to the large occupation fraction of MBHs at all times \citep{HM1993}. These systems have recently received considerable attention as their coalescence produces the loudest sources of gravitational waves (GWs) in the Universe \citep[][]{peters1964}. 

Depending on their masses, BHBs are expected to shine at GW frequencies between $10^{-9}$ Hz up to $10^{-1}$ Hz, a broad frequency window whose lower end is accessible by current observatories like the Pulsar Timing Array (PTA, \citealt{Desvignes2016, Reardon2016, Perera2019, Alam2021}) and the highest end by upcoming missions like the Laser Space Interferometer (LISA,  \citealt{Amaro-Seoane2017, Schodel2017, Barack2019}). Detection of GWs from BHBs would allow to build a statistical sample of key parameters like the masses, spins, orientations and even distances of MBHs. Given that BHBs can be seen even at very large red-shifts,  it could also provide an exciting new cosmological probe \citep[e.g.][]{V2021}. However, bridging the gap between the kpc-scale separation of the initial merger and the mpc-scale separation required for GW inspiral cannot be taken for granted and requires that binary hardening remains efficient over several orders of magnitudes in separation.

The evolution of a massive black hole binary (BHB) in a gas-poor system depends upon the properties of the merging galaxies as well as the orbital parameters. Qualitatively, the evolution can be divided in three separate phases \citep[][]{Begelman1980}: (i) a first phase driven by dynamical friction against the dark matter and stellar content of the galaxy that brings the MBHs together in a pair \citep{chandrasekhar1943}, if it is efficient enough;(ii) a second phase of rapid hardening driven by encounters with stars on low angular momentum orbits \citep{quinlan1996, sesana2006} and (iii) a third phase characterised by hardening against stars in the losscone\footnote{The `binary losscone' is defined as the subset of stars in the galaxy on low angular momentum orbits, that can strongly interact with the central BHB.} followed, if efficient, by GW emission and coalescence to a single MBH \citep{peters1964}.

%While the dynamical friction phase is estimated to be rapid for any realistic MBH masses (e.g. mass ratios $q> 10^{-3}$; \citealt{GM2007}) the hardening phase may slow down and even lead to stalling \citep{MM2001}.
The dynamical friction phase is estimated to be rapid for any realistic MBH masses (e.g. mass ratios $q> 10^{-3}$; \citealt{GM2007}), with only binaries with $q< 10^{-3}$ showing evidence of stalling \citep{DA2017}. The hardening phase, however, may be characterised by a slowing decay or even stalling \citep{MM2001}. This occurs in spherical stellar distributions where the losscone is emptied after approximately one dynamical time due to the ejection of all stars initially on centrophilic orbits. The only dynamical process able to refill the losscone in spherical potentials is two-body relaxation, but this operates on the relaxation timescale, which is longer than the age of the Universe for massive galaxies. However, it has been shown that even a mild triaxiality in the stellar system is sufficient to trigger collisionless losscone refilling leading to coalescence on timescales shorter than a Hubble time \citep{Berczik2006, VAM2015, G2017}.  This occurs because the total angular momentum of stellar orbits is not conserved in non-spherical potentials and torques lead to a replenishment of centrophilic orbits \citep{yu2002}. Rotation also enhances binary hardening \citep{holley+15, Mirza2017}. Triaxiality and/or rotation are naturally achieved in galactic mergers \citep{khan16, Bortolas2018}. 
It is therefore legitimate to conclude that BHBs formed in mergers involving massive elliptical galaxies will find their way to coalescence. This is in agreement with the small number of available observations of BHBs, despite targeted searches at different spatial scales \citep[see e.g.][for a review]{Dotti2012}.

With the PTA possibly making its first detection within the decade \citep{Taylor2019} and LISA in its final planning phase, merger timescales for BHBs from numerical simulations have become crucial to estimate detection rates for both probes. However, estimates are highly uncertain, as they depend critically on the eccentricity of the newly-formed BHB and the properties of the host galaxies, with timescales spanning the range from tens of Myr to several Gyr \citep{GM2012, Khan2013, Rantala2017, khan18a, khan18b}.

The evolution of binaries in gas poor systems past the hard binary separation can be satisfactorily modelled via semi-analytical prescriptions \citep{sesanakhan2015}. These tools well describe the binary evolution in an isotropic stellar background both in the hardening stage  \citep{Sesana2010} and along the GW-driven decay \citep[][]{peters1964}. Tracking the evolution of the orbital elements shows that while the semi-major axis $a$ shrinks in both phases, the eccentricity $e$ grows in the hardening stage (especially for initially non circular binaries with mass ratios $q=M_2/M_1\ll 1$, where $M_2\leq M_1$ by definition) but is then quickly dissipated in the GW-dominated regime. It is of primary importance to correctly predict the evolution of the eccentricity at the end of the hardening phase
as this sets both the onset and duration of the GW phase: GW emission begins earlier for eccentric binaries and the timescale to coalescence
scales as $\propto a^4 (1-e^2)^{7/2}$; but see also \cite{Zwick2020, Zwick2021}. 

However, self-consistently modelling the binary evolution from the large scale of the galaxy merger to the scale of binary formation and hardening and then coalescence is a computational challenge. To date, it remains unclear whether the initial parameters of the galaxy merger impact the eccentricity at the binary formation and then at the onset of the hardening stage, thus potentially significantly impacting the evolutionary timescale of binaries. 

Among all phases of binary evolution, the most uncertain is the phase of binary formation, where the BHB transitions from a dynamical friction dominated evolution to a stellar slingshot dominated evolution, partly due to the difficulty of modelling the effects of stochasticity \citep{nasim2020}.

In this study, we combine state-of-the-art direct summation $N$-body simulations of galaxy mergers with semi-analytical models to investigate the process of binary formation and self-consistently constrain the eccentricity of BHBs from large to small scales.
We find that the binding of BHs does not happen instantaneously but rather it is a process during which the BHs oscillate between a bound and an unbound state several times before settling into a bound two-body Keplerian orbit. 
The process is intrinsically chaotic and, depending on the orbital parameters of the merger, the mass ratio and the properties of the galaxy, results in different orbital elements for the newly formed BHB. 
The main effect of the chaotic phase of binding is to introduce a scatter in  the eccentricity of BHBs at the time of binary formation.
However, we identify a tight correlation between the eccentricity at binding and the properties of the merger, namely the initial orbital eccentricity. In addition, we find a clear correlation between the eccentricity at binding and that at the onset of the hardening phase, which can be exploited to tailor semi-analytic predictions of timescales for GW coalescence.

\section{Numerical methods}

\begin{table*}
\begin{center}
\caption{Galaxy models: identifier, slope of the initial density profile, initial orbital eccentricity of the merger, galaxy mass ratio, semi-major axis as defined by Eq.\,\ref{eq:af}, semi-major axis at binding, hard-binary separation as defined by Eq.\,\ref{eq:ah}, time of black hole binding, time when the hard-binary separation is reached, eccentricity at the time of binding, eccentricity at the hard-binary separation.}
\label{tab:data}
\begin{tabular}{lcccccccccc} 
\hline
Name & $\gamma$ & $e$ & $q$ & $a_f$ & $a_b$ & $a_h$ & $t_b$ & $t_h$ & $e_b$ & $e_h$\\
\hline
G05E05Q1   & 0.5 & 0.5 & 1 & 0.146 & 0.047 & 0.0093 & 351 & 371 & 0.22 & 0.065\\
G05E05Q025 & 0.5 & 0.5 & 0.25 & 0.095 & 0.048 & 0.0034 & 954 & 1010 & 0.135 & 0.14\\
G05E05Q01  & 0.5 & 0.5 & 0.1 & - & - & -& - & -& -&-\\
\hline
G05E07Q1   & 0.5 & 0.7 & 1 & 0.157 & 0.026 & 0.009 & 238.5 & 257 & 0.815 &  0.64\\
G05E07Q025 & 0.5 & 0.7 & 0.25 & 0.096 & 0.042 & 0.0035 & 757 & 825.25 & 0.26  & 0.10\\
G05E07Q01  & 0.5 & 0.7 & 0.1 & - & - & -& - & -& -&-\\
\hline
G05E09Q1   & 0.5 & 0.9 & 1 & 0.132 & 0.037 & 0.008 & 140 &  159& 0.625 & 0.72\\
G05E09Q025 & 0.5 & 0.9 & 0.25 & 0.101& 0.033 & 0.0033 & 364 & 436 & 0.45 & 0.31\\
G05E09Q01  & 0.5 & 0.9 & 0.1 & 0.060 & 0.039 & 0.0012  & 2168 & 2337 & 0.52 & 0.80\\
\hline
G1E05Q1   & 1 & 0.5 & 1 & 0.093 & 0.026 & 0.006 & 370.5 & 375.5 & 0.16 & 0.08\\
G1E05Q025 & 1 & 0.5 & 0.25 & 0.057 & 0.0022  & 0.0021 & 745 & 780 & 0.23 & 0.09\\
G1E05Q01  & 1.0 & 0.5 & 0.1 & - & - & -& - & -& -&-\\
\hline
G1E07Q1   & 1 & 0.7 & 1 & 0.0496 & 0.0274 & 0.0050& 242.75 & 249.75 & 0.39 &  0.21\\
G1E07Q025 & 1 & 0.7 & 0.25 & 0.0375 & 0.0249 & 0.00189 &  569.25 & 607.25 & 0.48 & 0.03\\
G1E07Q01  & 1 & 0.7 & 0.1 & 0.030 & 0.0187 & 0.0009 & 3910.75 & 3996.75 & 0.16 & 0.38\\
\hline
G1E09Q1   & 1 & 0.9 & 1 & 0.0603 & 0.012 & 0.0052 & 127 & 133 & 0.56 & 0.58\\
G1E09Q025 & 1 & 0.9 & 0.25 & 0.048 & 0.019 & 0.0023 & 289.5 & 316 & 0.45& 0.21\\
G1E09Q01  & 1 & 0.9 & 0.1 & 0.028 & 0.024 & 0.0008 & 1573 & 1639.75 & 0.34 & 0.63 \\
\hline
G15E05Q1   & 1.5 & 0.5 & 1 & 0.043 & 0.006 & 0.0027 & 409 & 411.5 & 0.115 & 0.045  \\
G15E05Q025 & 1.5 & 0.5 & 0.25 & 0.018 & 0.0077 & 0.0010 & 648 & 665.5 & 0.0385 & 0.045\\
G15E05Q01  & 1.5 & 0.5 & 0.1 & 0.009 & 0.0036 & 0.00058 & 2098.5 & 2138 & 0.19 & 0.24\\
\hline
G15E07Q1   & 1.5 & 0.7 & 1 & 0.040 & 0.011 & 0.00133 & 256.5 & 265.5 & 0.11 & 0.12\\
G15E07Q025 & 1.5 & 0.7 & 0.25 & 0.018 & 0.0094 & 0.00035 & 455.5& 548 & 0.0077 & 0.155\\
G15E07Q01  & 1.5 & 0.7 & 0.1 & - & - & -& - & -& -&-\\
\hline
G15E09Q1   & 1.5 & 0.9 & 1 & 0.0165 & 0.0065 & 0.00039 & 119.5 & 180 & 0.37 & 0.52\\
G15E09Q025 & 1.5 & 0.9 & 0.25 & 0.022 & 0.0042 & 0.00044 & 220.5 & 280 & 0.18& 0.24\\
G15E09Q01  & 1.5 & 0.9 & 0.1 & 0.019 & 0.0114 & 0.00057 & 1325.5 & 1355 & 0.26& 0.41\\

\hline
\end{tabular} 
\end{center}
\end{table*}

We model the merger of two spherical non-rotating galaxies with the direct summation code $\phi$-GPU \citep{berczik+11, Just2012, sobolenko2017, 2019MNRAS.484.3279P}, a fourth order Hermite integrator supporting the accelerated computation of gravitational forces on Graphic Processing Units. $\phi$-GPU employs gravitational softening ($\epsilon$) for the calculation of gravitational forces among all particles. In order to achieve the most accurate integration of the binary orbit, we set $\epsilon = 0$ for MBH-MBH interactions and $\epsilon = 7 \times 10^{-6}$ for MBH-star interactions. We allow a small softening $\epsilon = 10^{-4}$ for star-star interactions to avoid the formation of stellar binaries.
The galaxies follow Dehnen's spherical density profile
\citep{Dehnen1993}:
\begin{equation}
\rho(r)=\frac{(3-\gamma) \,M}{4\pi}\frac{r_0}{r^\gamma(r+r_0)^{4-\gamma}},
\end{equation}
where $M_{\rm gal}$ is the total mass of the galaxy, $r_0$ is the scale radius of the model and $\gamma$ is the central density slope.  We adopt three values of $\gamma=0.5, 1.0, 1.5$ to model shallow and cuspy radial density profiles. 
The galaxies are placed on bound Keplerian orbits of different eccentricities $e=0.5, 0.7, 0.9$ at a fixed distance of $D=20\,r_0$ and semi-major axis $a=15r_0$\footnote{The radial and tangential components of the relative velocity are $\vrad=-0.072$, $\vtan=0.168$ for $e = 0.5$, $\vrad = -0.119$
$\vtan = 0.138$ for $e=0.7$, $\vrad = -0.162$, $\vtan = 0.084$ for $e=0.9$, in normalised code units.}. We consider both major and minor mergers with mass ratio $q=1.0, 0.25, 0.1$. All galaxies host a central MBH of fixed mass $\mbh=0.005$ in units of the galaxy mass. This is chosen as a typical value for massive elliptical galaxies, though we note that a large spread is observed ($0.0005 \lesssim q \lesssim 0.01$; \citealp{Schutte2019}) and that often models like ours represent the inner rather than the whole bulge of the galaxy. This implies that the interacting MBHs have the same mass ratio $q$ as their host galaxies. The merging galaxies have an equal scale radius $r_0$ regardless of $q$. 
Even though the scale radius is often scaled with the mass, this choice is not representative of the huge variety of different galaxies; for instance \citet[][]{Emami2021} show that, for their subclass of dwarfs, the radius scales very weakly with the stellar mass. We warn that this choice makes the infalling galaxy more susceptible to tidal effects and stripping, which may affect our findings.

The parameters of the merging galaxies are listed in Table \ref{tab:data} with the model identifiers that will be used throughout. These include the slope of the density profile (label 'G'), the initial orbital eccentricity of the merger (label 'E') and the merger mass ratio (label 'Q'). All models are generated at a resolution of $N=512$k particles. Additional realisations are generated for model G05E07Q1 (5 independent realisations) and at higher resolution ($N=1M$) to investigate the dependence of our results on particle number and random noise (see section \ref{sec:noise}). 

\begin{table}
\begin{center}
\caption{Scaling to physical units [L] (distance), [T] (time) and [V] (velocity) for two adopted values of the binary mass as a function of the central slope of the merging galaxies. Here $M_{\rm gal}$ has to be intended as the total mass of the galaxy merger remnant, which is set equal to 1 in internal units.}
\label{tab:units}
\begin{tabular}{ccccc} 
\hline 
$M_b, M_{\rm gal}$(M$_\odot$) & $\gamma$ & [L] (pc) & [T] (Myr) & [V] (km s$^-1$)\\
\hline
$4\times10^6$, $8\times10^8$  & 0.5 & 31 & 0.089 & 335\\
$4\times10^6$, $8\times10^8$  & 1.0 & 52 & 0.20 & 257 \\
$4\times10^6$, $8\times10^8$  & 1.5 & 119 & 0.68 & 170\\
\hline
$1\times10^8$, $2\times10^{10}$ & 0.5 & 186 & 0.27 & 680\\
$1\times10^8$, $2\times10^{10}$ & 1.0 & 315 & 0.59 & 523\\
$1\times10^8$, $2\times10^{10}$ & 1.5 & 719 & 2.0  & 346\\
\hline
\end{tabular} 
\end{center}
\end{table}
We adopt normalised code units such that the Newtonian gravitational constant, the scale radius of the density profile and the total mass of the merging galaxies are set equal to unity: $G=1$, $r_0=1$ and $M_{\rm tot}=1$. Scaling to physical units can be performed by equating the value of the central MBH's influence radius measured in the initial model with the value obtained by \citet{merritt2009}, $r_{\rm infl} = 35 \,{\rm pc}\left( M_{\bullet}/10^8 \msun\right)^{0.56}$.
Length, time and velocity units are given in Table \ref{tab:units} for a Milky Way type galaxy with a MBH of mass $4.0\times10^6\msun$ and a generic one with a MBH of mass $10^8\msun$.

Different models show binary evolution on different timescales, therefore rather than fixing a total integration time we monitor each model and extend the time integration well past binary formation and, in most instances, past the formation of a hard binary. Some of the minor merger models evolve so slowly in the dynamical friction phase that it wasn't possible to reach binary formation; we exclude those models from our analysis (see Table \ref{tab:data}).

\section{Evolution of massive black hole binaries}
\label{sec:evol}
We consider binaries of supermassive black holes of masses $M_1>M_2$, total mass $M_b$ and a mass ratio $q=M_2/M_1$. Their evolution shows three characteristic phases during which the hardening is driven by different processes. In the first phase, dynamical friction against the stellar population drags the MBHs to a separation $a_f$, defined as the separation where the enclosed stellar mass is of the order of twice the mass $M_2$ of the secondary
\begin{equation}
\label{eq:af}
M_*(<a_f) = 2M_2.
\end{equation}
In the case of equal mass mergers, this is equivalent to the separation enclosing a mass equal to the binary mass $M_b$. 
The time at which this separation is reached $t_f = t(a_f)$ is often taken to mark the time of binary formation, when the MBHs become bound. In fact, as we show in section \ref{sec:binary}, binary formation is not yet complete. More appropriately, $t_f$ roughly corresponds to the end of the merger process and can be taken to mark the end of the dynamical friction phase. Beyond this time, dynamical friction becomes less and less efficient while encounters with stars begin to remove energy and angular momentum from the pair. The second phase of evolution is a phase of very rapid and efficient hardening driven by three-body encounters between the binary and stars on intersecting orbits. Stars initially on losscone orbits are ejected by the binary on roughly a dynamical timescale, leading to a reduction in stellar density and the formation of a core. This process of core scouring is considered responsible for the observation of large cores in massive elliptical galaxies \citep[e.g][]{MM2001,merritt2006, GM2012}.

The binary is considered {\it hard} when it reaches the separation
\begin{equation}
\label{eq:ah}
    a_h = \frac{GM_2}{4\sigma^2}
\end{equation}
called the hard-binary separation, where its binding energy per unit mass exceeds the kinetic energy per unit mass of the stellar population, with $\sigma$ the stellar velocity dispersion \citep{MM2001}. 
The characteristic separations $a_f$ and $a_h$ depend on the slope of the density profile $\gamma$, with steeper profiles resulting in smaller separations \citep{Bortolas2018}. 
The time $t_h = t(a_h)$  at which the hard-binary separation is reached approximately marks the end of the hardening phase, as by this time all stars on initial losscone orbits have been ejected by the binary. The evolution into the third phase depends on the efficiency of losscone repopulation, which is determined by the geometry and kinematics of the nucleus. Stalling of the binary at the hard-binary separation (of the order of $1\pc$ for typical galaxy masses) is seen in simulations of BHBs evolving in non-rotating spherical potentials, due to the slow refilling of losscone orbits by two-body relaxation; a process often called {\it The Final Parsec Problem} \citep{milos2002}. Simulations of binaries in realistic merger remnants, which are characterised by triaxial shapes and/or some degree of rotation, show continuing evolution and sustained hardening at rates sufficient to lead to GW emission and coalescence in much less than a Hubble time \citep{preto2011, khan2011, VAM2015, G2017, Arca2019}.

\section{Results}
\label{sec:results}
\begin{figure}
    \centering
    \includegraphics[width=1.1\columnwidth]{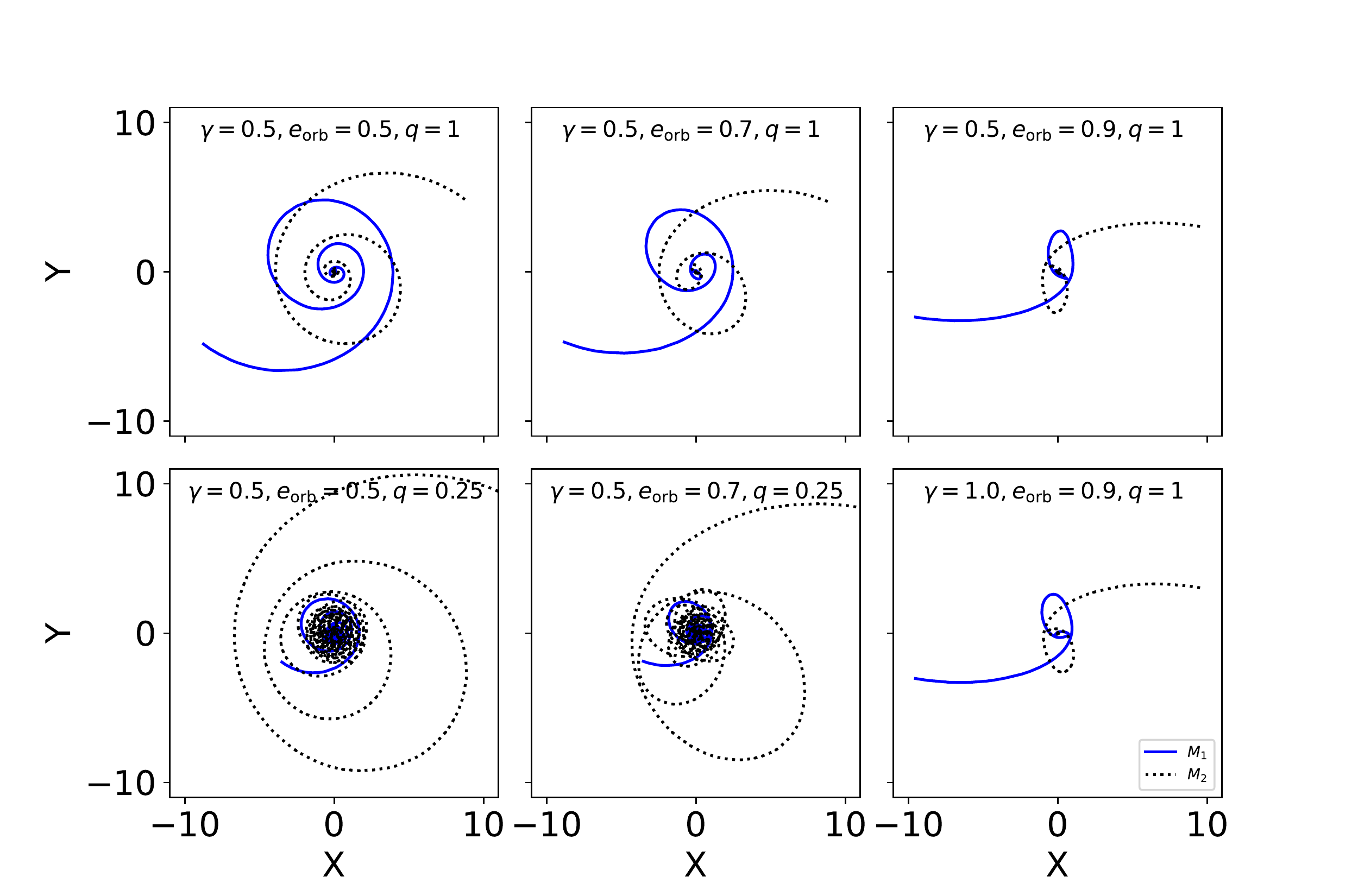}    
    \caption{Trajectories of the MBHs in 6 representative models with different initial orbital eccentricity, mass ratio and density slope: primary MBH (solid line) and secondary MBH (dotted line).  The initial orbit of the galaxy pair lies in the $z = 0$ plane.}
    \label{fig:orbits}
\end{figure}

The large scale evolution of the mergers depends mainly on the initial orbital eccentricity, with more eccentric orbits resulting in fewer pericentre passages before merging, and on the mass ratio. The trajectories of the MBHs trace the large scale orbits of the host galaxies, as shown in Fig.\,\ref{fig:orbits}, and do not show any significant dependence on the slope $\gamma$ of the central density profile. 

\begin{figure*}
    \centering
    \includegraphics[width=1.0\columnwidth]{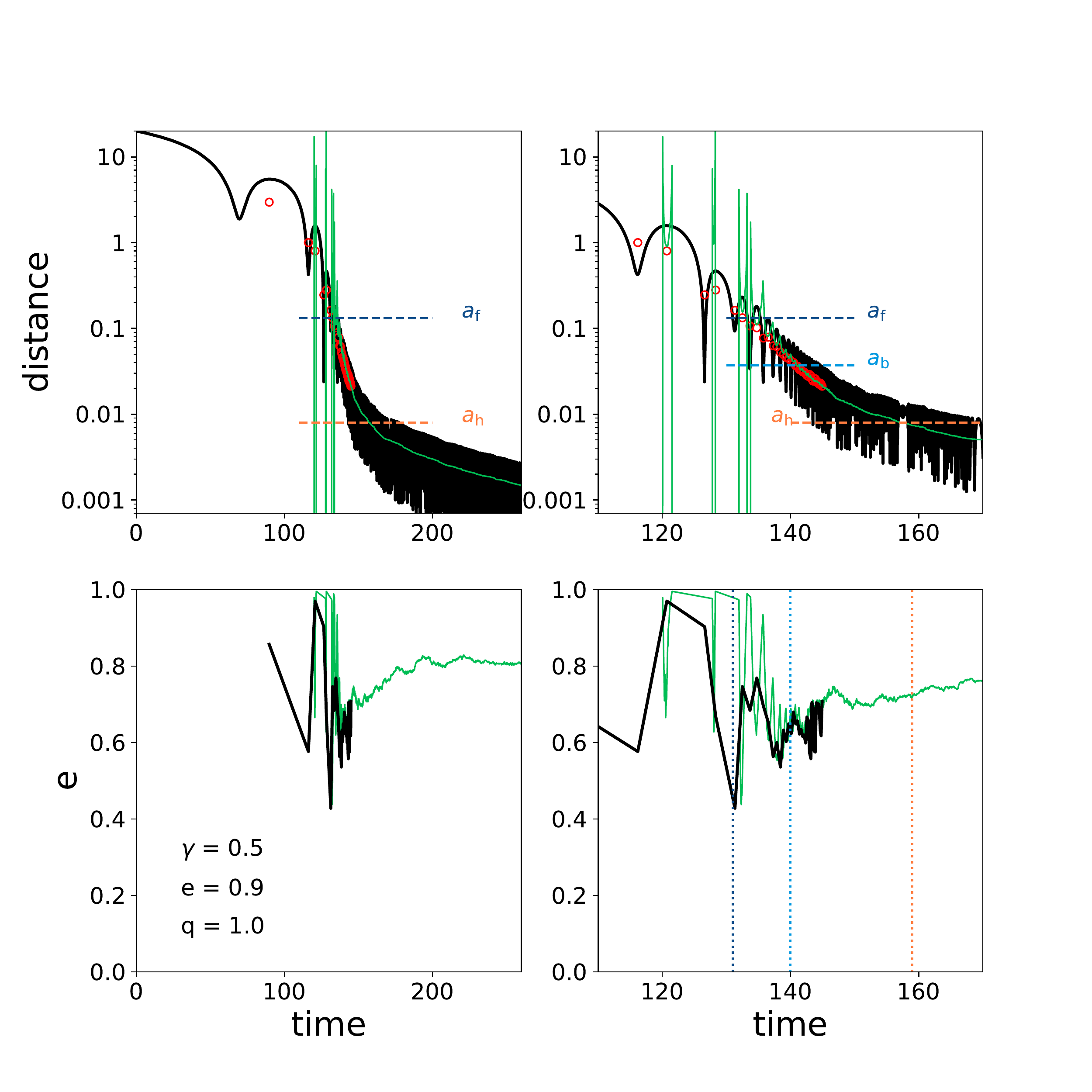}
    \includegraphics[width=1.0\columnwidth]{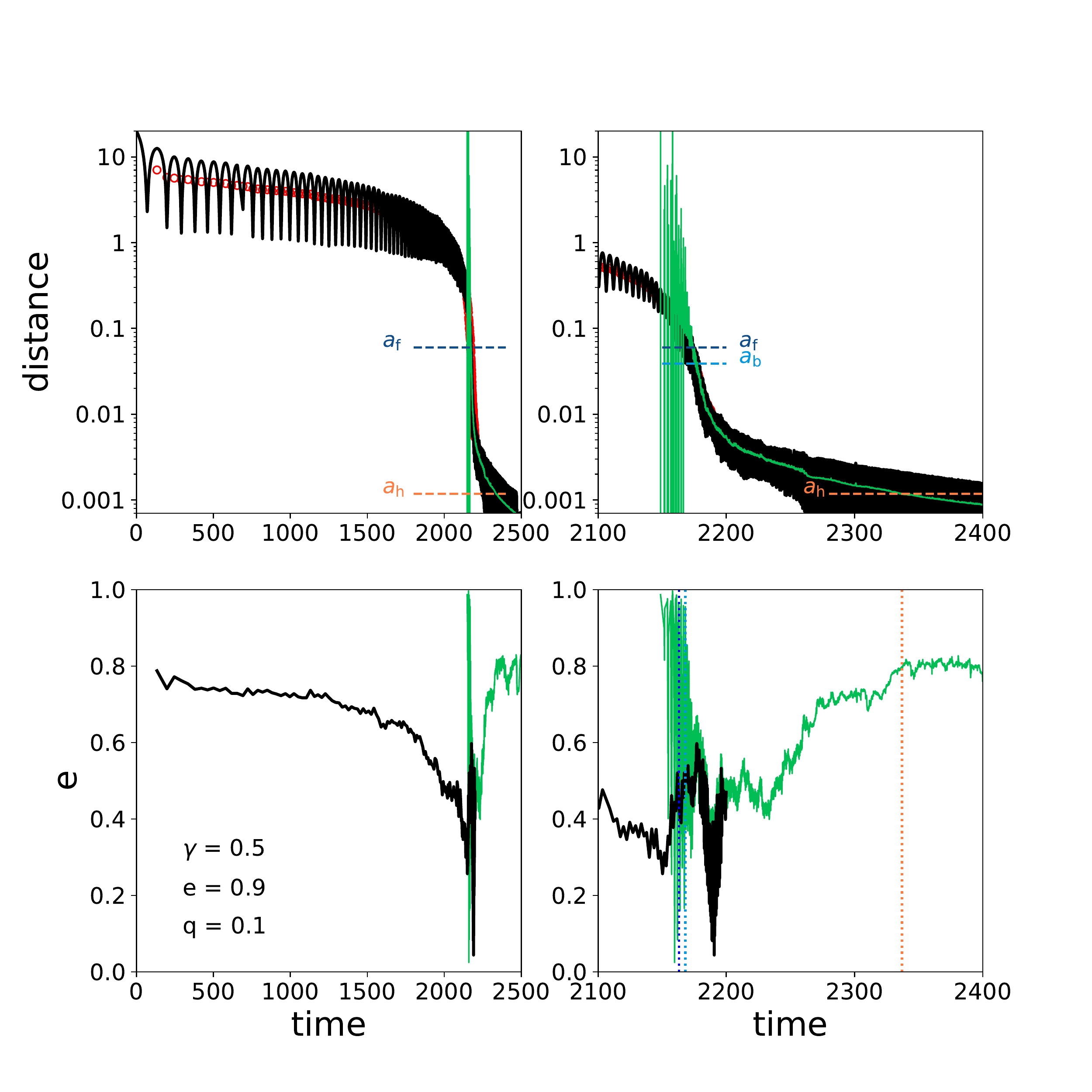}
    \caption{(Top left) Evolution of the distance between the MBHs (black solid line) and of the Keplerian semi-major axis (green solid line) as a function of time for models. The characteristic separations $a_f$ and $a_h$ are shown with horizontal dashed lines. The red points mark the semi-major axis computed via subsequent pericentres and apocentres. (Top right) A zoomed-in version of the top left panel to illustrate the process of MBH binding, showing also the separation $a_b$ which we take to mark binary formation. (Bottom left) Evolution of the orbital eccentricity of the BHB (black solid line) computed numerically from the pericentre and apocentre of the orbit and of the Keplerian eccentricity (green solid line). (Bottom right)  A zoomed-in version of the bottom left panel, marking the times when the critical separations are reached: $t_f$, $t_b$ and $t_h$. The left-hand figure refers to model G05E09Q1, while the right-hand one is for model G05E09Q01.}
    \label{fig:dist1}
\end{figure*}

\begin{figure*}
    \centering
    \includegraphics[width=1.0\columnwidth]{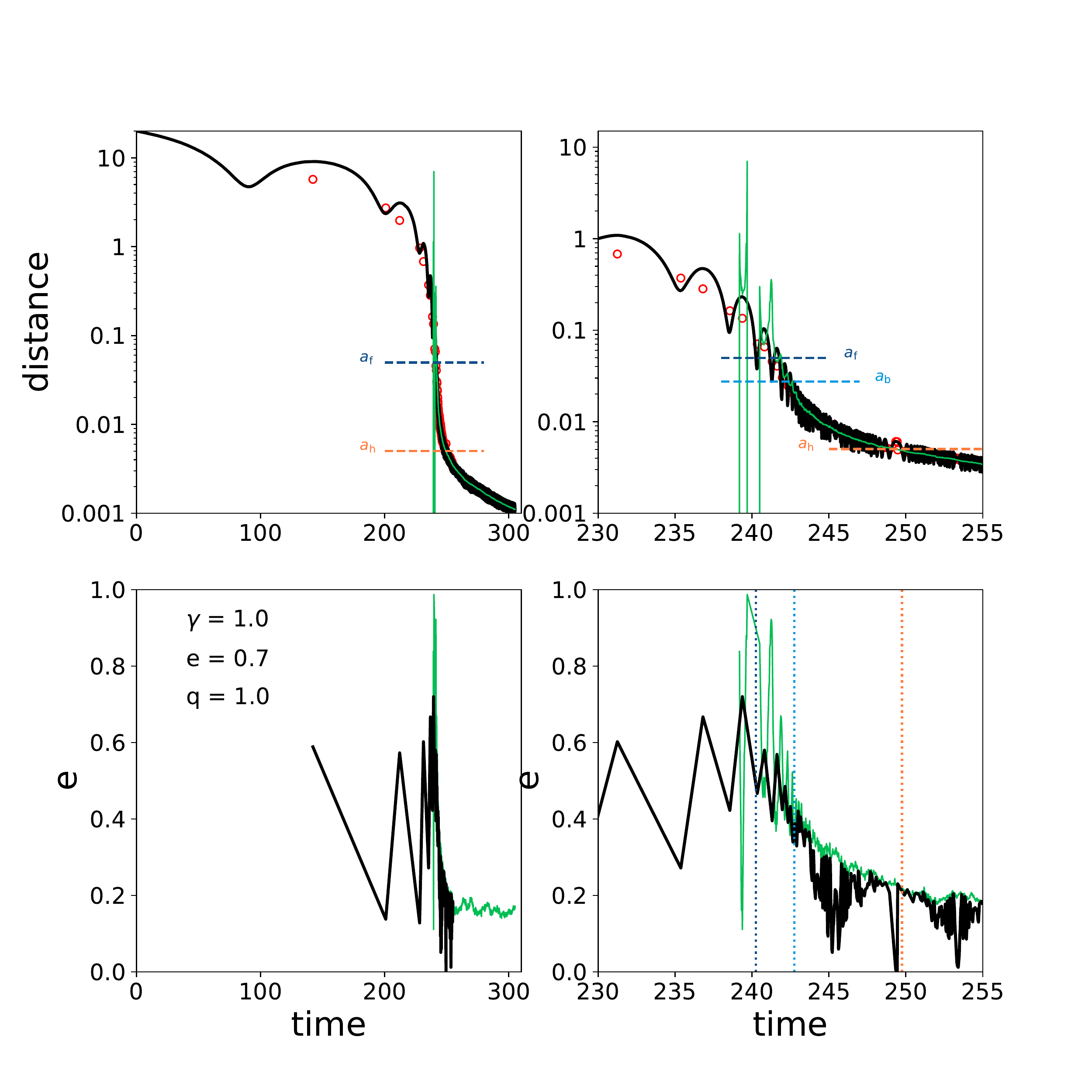}
    \includegraphics[width=1.0\columnwidth]{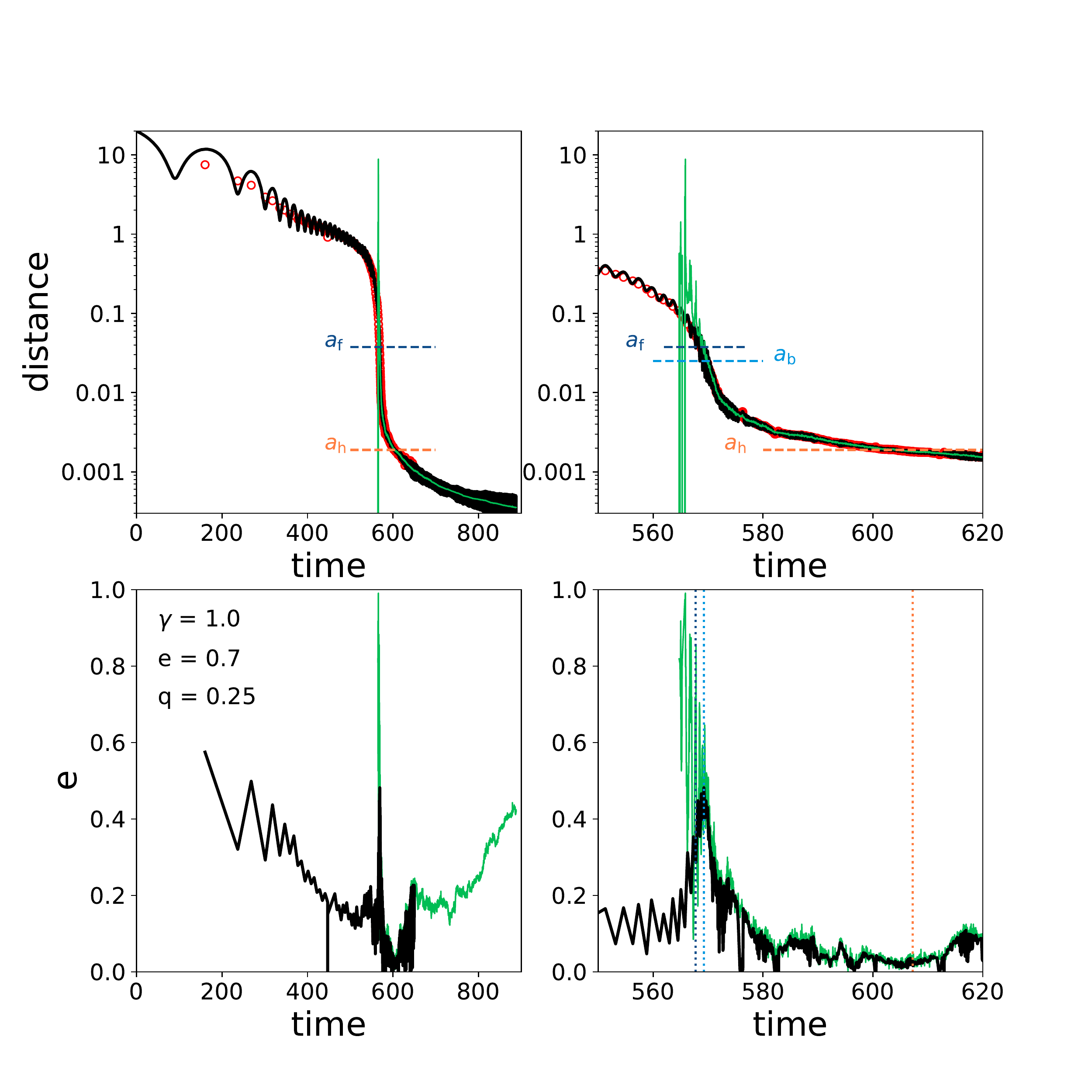}
    \caption{Like Fig.\,\ref{fig:dist1} for models G1E07Q1 and G1E07Q025.}
    \label{fig:dist2}
\end{figure*}

\begin{figure*}
    \centering
    \includegraphics[width=1.0\columnwidth]{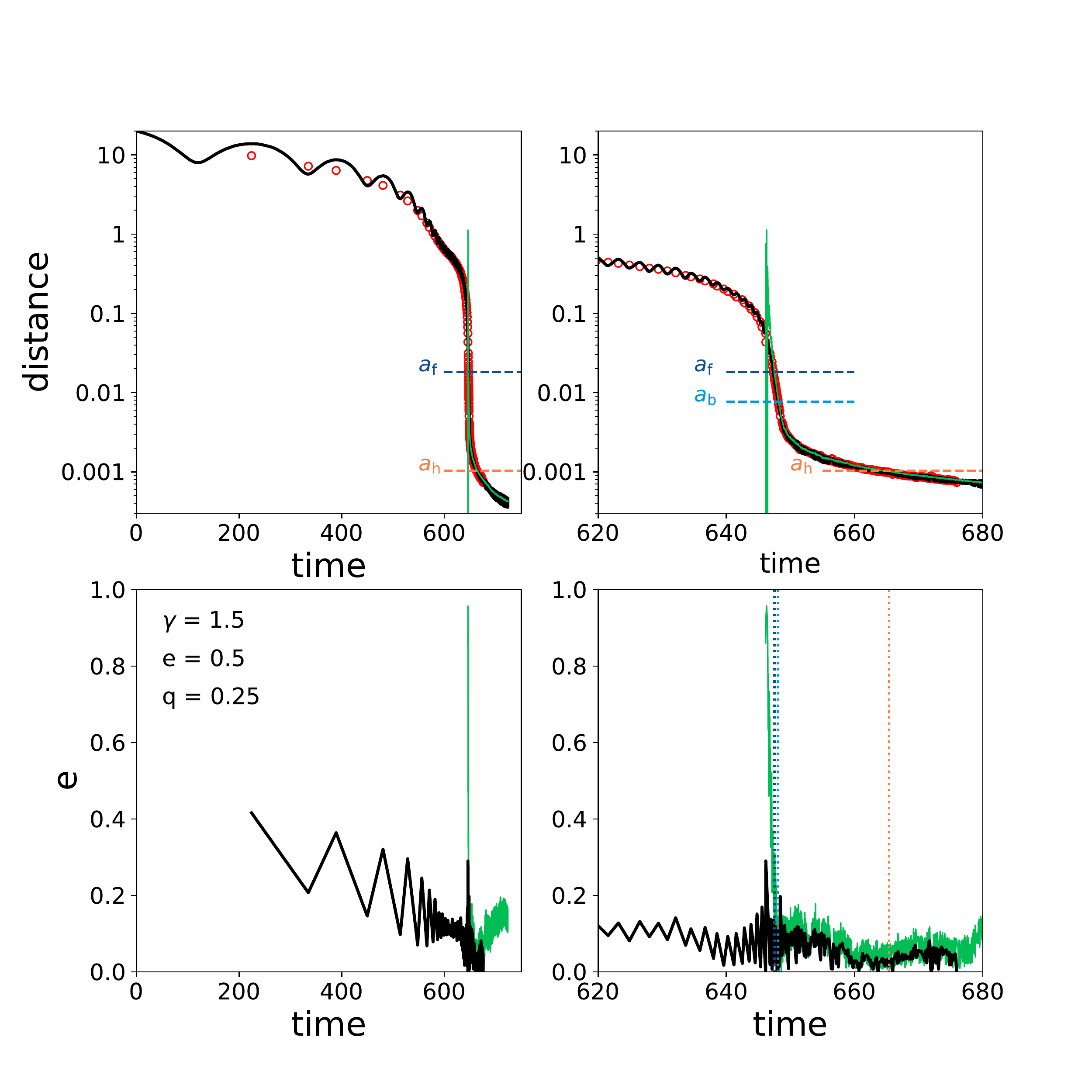}
    \includegraphics[width=1.0\columnwidth]{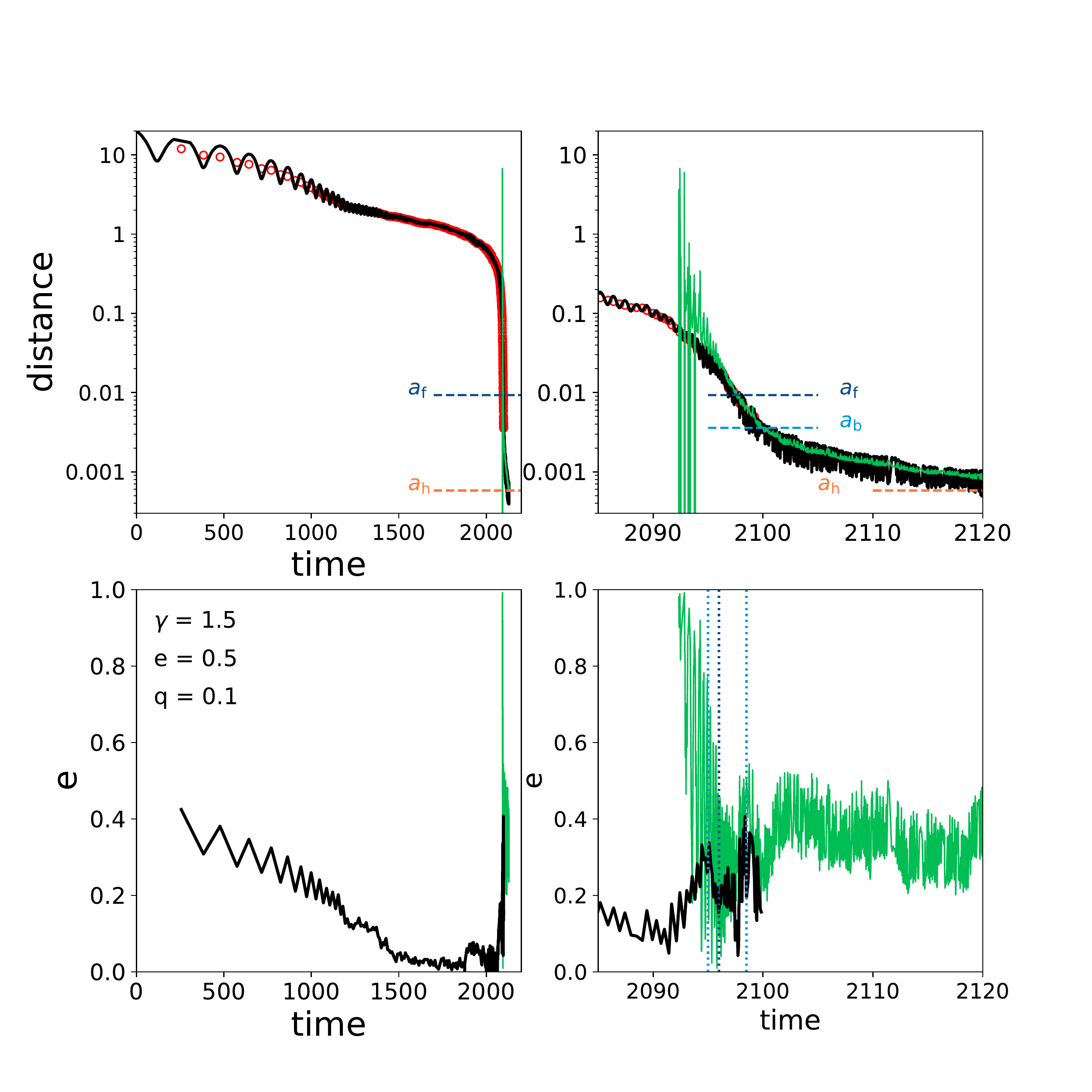}
     \caption{Like Fig.\,\ref{fig:dist1} for models G15E05Q025 and G15E05Q01.}
   \label{fig:dist3}
\end{figure*}

The evolution of the MBHs from large to small scales shows the three distinct phases described in section \ref{sec:evol}, as can be seen in the top panels of Fig.\,\ref{fig:dist1}, \ref{fig:dist2}, \ref{fig:dist3} for some representative models. 

We compute the characteristic separation $a_f$ which marks the end of the dynamical friction phase as well as the hard-binary separation $a_h$ for all models. Values are given in Table \,\ref{tab:data} and are marked in Fig.\,\ref{fig:dist1}, \ref{fig:dist2}, \ref{fig:dist3} for a selection of models.

\subsection{Binary formation}
\label{sec:binary}
In this section we aim to investigate the phase of binary formation. This phase is crucial as it sets the early eccentricity of BHBs which then evolve through hardening towards GW emission and coalescence. As we show below, it is important to properly define the time of binary formation, prior to which the eccentricity oscillates strongly, in order to accurately track the eccentricity evolution in the subsequent stages, with strong implications on merger timescales and GW detection rates.

Figures \ref{fig:dist1}, \ref{fig:dist2}, \ref{fig:dist3}, 
show the evolution of the distance between the MBHs as a function of time, together with the semi-major axis (top panels) and the evolution of the eccentricity (bottom panels). The Keplerian orbital elements are well-defined only when the binary is bound. To monitor the evolution of the semi-major axis and the eccentricity at early times, when the binary is not yet bound, we track subsequent pericentre ($r_p$) and apocentre ($r_a$) passages and compute the orbital semi-major and eccentricity as:
\begin{align}\label{eq:peri_apo_params}
    e = \frac{r_a-r_p}{r_a+r_p}\\
    a = \frac{r_a+r_p}{2}\,.
\end{align}
We note that the Keplerian orbital elements are characterised by large oscillations around the time of binary formation, suggesting that binary binding is a phase rather than an instantaneous process. Despite being affected by stochastic effects, we observe some periodicity in the oscillations, with large variations found around apocentre, when the MBHs are further apart and more stellar mass is enclosed within the orbit. Furthermore, the Keplerian elements, which are meant to describe an isolated two-body system, are inadequate at describing the behaviour of the MBHs in the rapidly evolving background of the galaxy merger, especially near apocentre. The orbital eccentricity is a useful quantity because it can be computed at any time, even when the MBHs are still widely separated, and matches the Keplerian eccentricity after binding. It becomes unreliable at late times, due to the difficulty in identifying pericentre and apocentre passages as the BHB orbital period quickly becomes very short and the output frequency in the simulation does not allow to properly resolve the time of pericentre and apocentre.  We therefore show both estimates of the eccentricity, and adopt the orbital eccentricity to describe the early evolution of the binary (up to binding) and the Keplerian eccentricity to describe the late evolution (after binding).

We now seek to establish a physically motivated criterion for binary formation. Given the large oscillations observed in the Keplerian elements, and the fact that the binary energy oscillates between positive and negative values multiple times before becoming permanently negative, it is incorrect to adopt the time the binding energy first becomes negative as the time of binary formation. One possibility is the time $t_f$ when the separation $a_f$ is reached, approximately marking the end of the dynamical friction phase. The figures however show that the MBHs are not yet bound, and the orbital elements still suffer large oscillations due to the chaotic nature of interactions with background stars. We therefore consider alternative separations enclosing different fractions of the total binary mass $M_b$. We find that the separation $a_b$ such that
\begin{equation}
\label{eq:bound}
    M(<a_b) = 0.1 M_b
\end{equation}
and the corresponding time $t_b$ provide an adequate representation of the time of binary formation for all models, as by this time most of the oscillations have settled and the system can be considered in a stable bound configuration. 
We define as eccentricity at binary formation $e_b$ the value of the orbital eccentricity at time $t_b$. Values of $a_b$, $e_b$ and $t_b$ are given in Table\,\ref{tab:data}. We note that considering the binary bound at an earlier time, for example at time $t_f$ or even at the first time the binding energy becomes negative \citep[as in e.g][]{nasim2021}, results in a larger value of the eccentricity attributed to the binary. This is because the eccentricity is, by definition, extremely large when the MBHs first pair together. However, the binary is not yet in a stable configuration, and the orbital elements suffer large variations during the binding phase.

\begin{figure*}
    \centering
    \includegraphics[width=2.2\columnwidth]{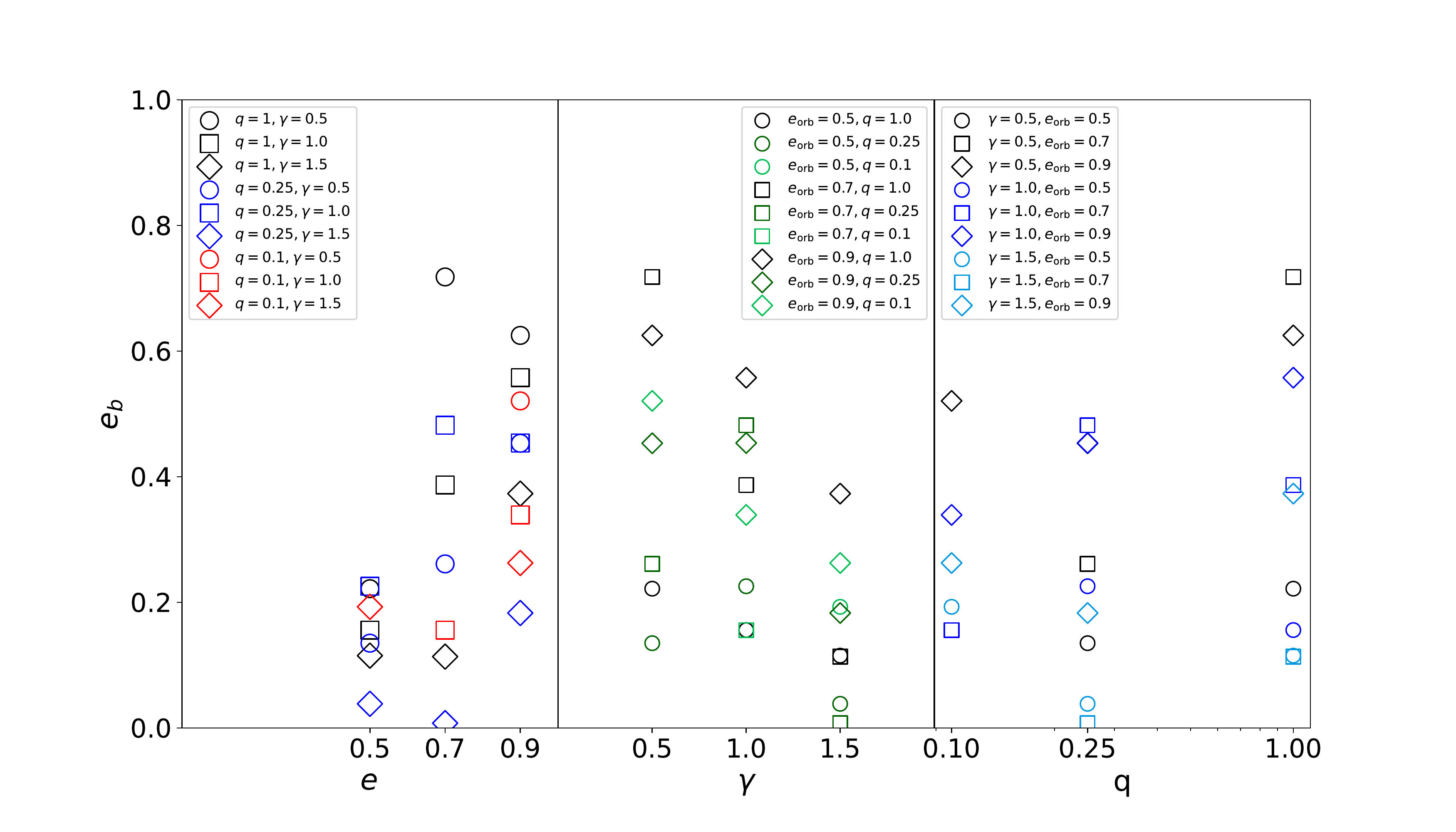}    
    \caption{Bound eccentricity $e_b$ as defined by Eq.\,\ref{eq:bound} as a function of (left) the initial orbital eccentricity $e$ of the merger;(middle) the central slope of the galaxy's density profile; (right) the MBH mass ratio $q$.}
    \label{fig:ebound}
\end{figure*}

\subsection{Eccentricity evolution}
Figure \ref{fig:ebound} shows the dependence of the bound eccentricity $e_b$ on the three parameters of the mergers: the orbital eccentricity $e$, the central slope $\gamma$ of the host galaxy's density profile and the mass ratio $q$ of the MBHs/galaxies. There is a clear correlation between the bound eccentricity and the initial eccentricity of the galactic orbit, with more radial orbits resulting in larger $e_b$, though with a considerable scatter.   We also note a weak dependence on the inner slope of the galaxy's profile, with steeper cusps resulting in less eccentric binaries at formation.  This can be understood in terms of increased circularization in steeper profiles in the dynamical friction phase, as explained below.
It is more difficult to draw conclusions on correlations with the mass ratio, as some $q=0.1$ models were not completed, but there is a suggested trend towards a smaller eccentricity when transitioning from equal mass mergers to 1:4 mergers, followed by a modest increase in eccentricity for 1:10 minor mergers. 

\begin{figure}
    \centering
    \includegraphics[width=1.0\columnwidth]{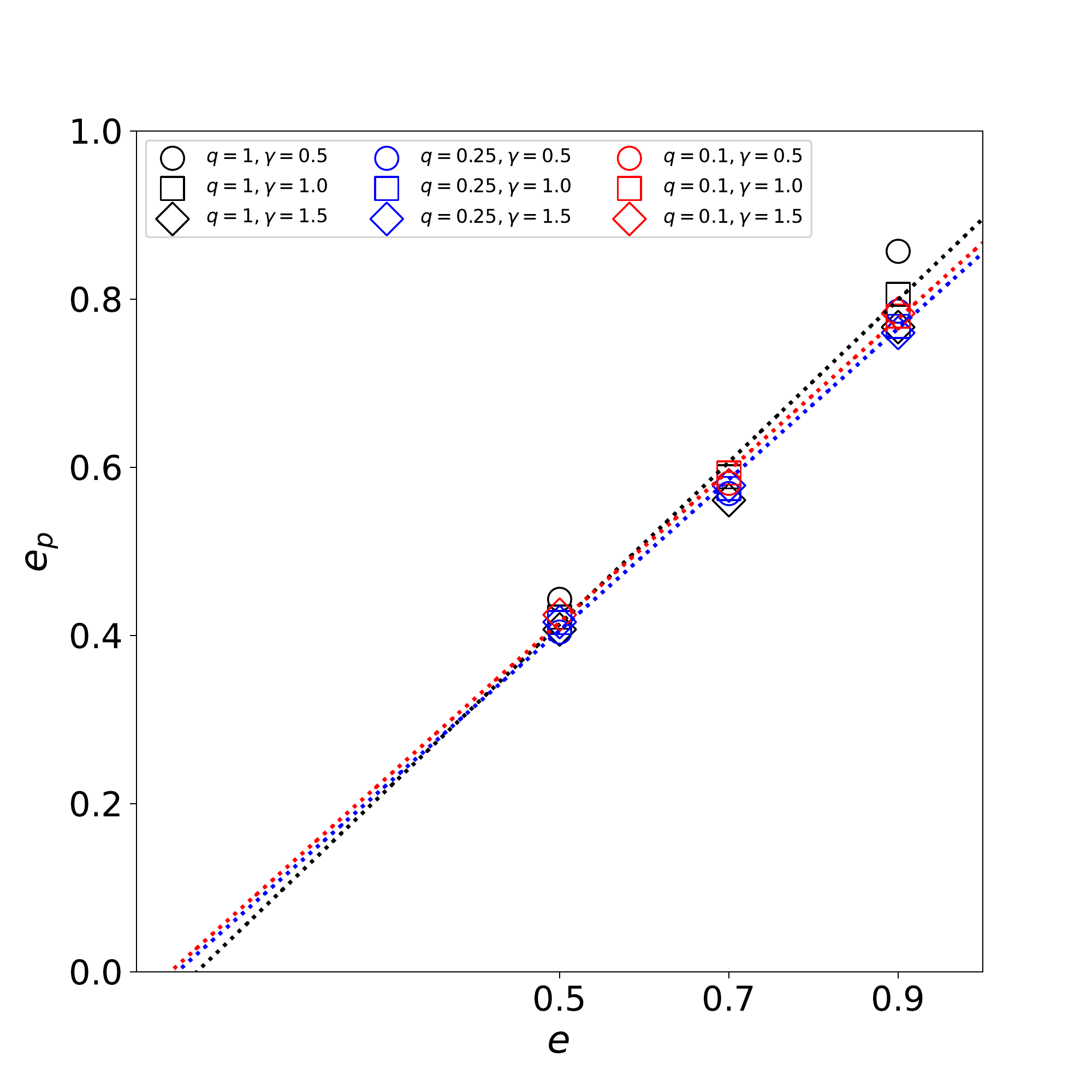}
    \caption{Orbital eccentricity of the MBHs computed at the first pericentre passage as a function of the initial eccentricity of the galactic merger, showing a strong correlation. The dotted lines represent the best linear regressions for models with different mass ratios: $q=1$ (black), $q=0.25$ (blue) and $q=0.1$ (red).}
    \label{fig:eperi}
\end{figure}
Interestingly, we see evidence of circularization in all models, which we attribute to the effects of the dynamical friction phase. In fact, the eccentricity during the early phases of the merger correlates with the eccentricity of the orbit as can be seen in Fig.\,\ref{fig:eperi}. This shows a strong correlation of the eccentricity at the first pericentre passage with the eccentricity of the galactic orbit.
\begin{figure}
    \centering
   \includegraphics[width=1.0\columnwidth]{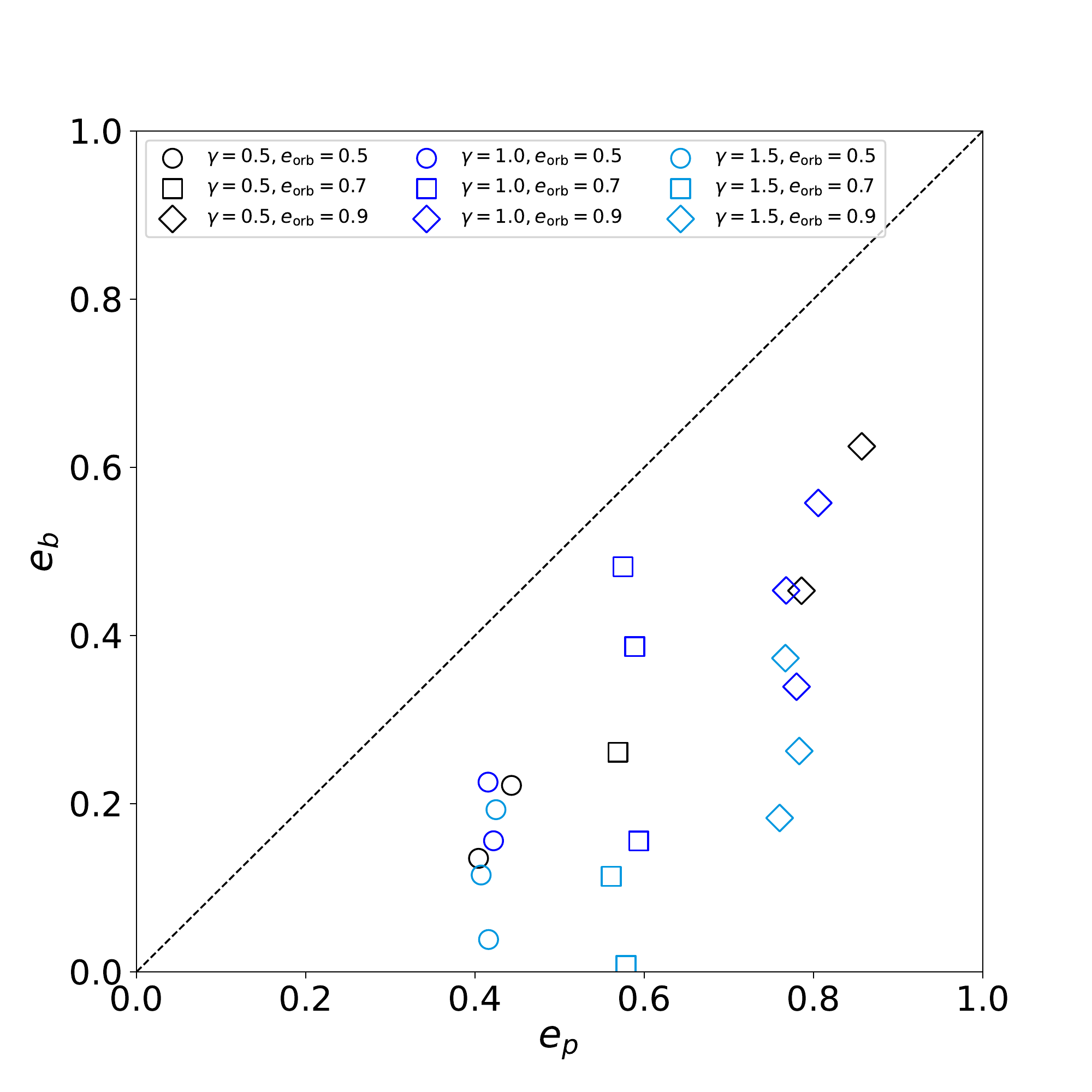}
    \caption{Relation between the eccentricity of the binary at binding, as defined by Eq.\,\ref{eq:bound}, and the eccentricity at the first pericentre passage during the galactic merger, showing evidence of circularization during the dynamical friction phase.}
    \label{fig:ebep}
\end{figure}
On the other hand, the bound eccentricity, which is computed after the dynamical friction phase, is significantly and systematically lower than that at the first pericentre passage, as shown in Fig.\,\ref{fig:ebep}.

\begin{figure}
    \centering
    \includegraphics[width=1.0\columnwidth]{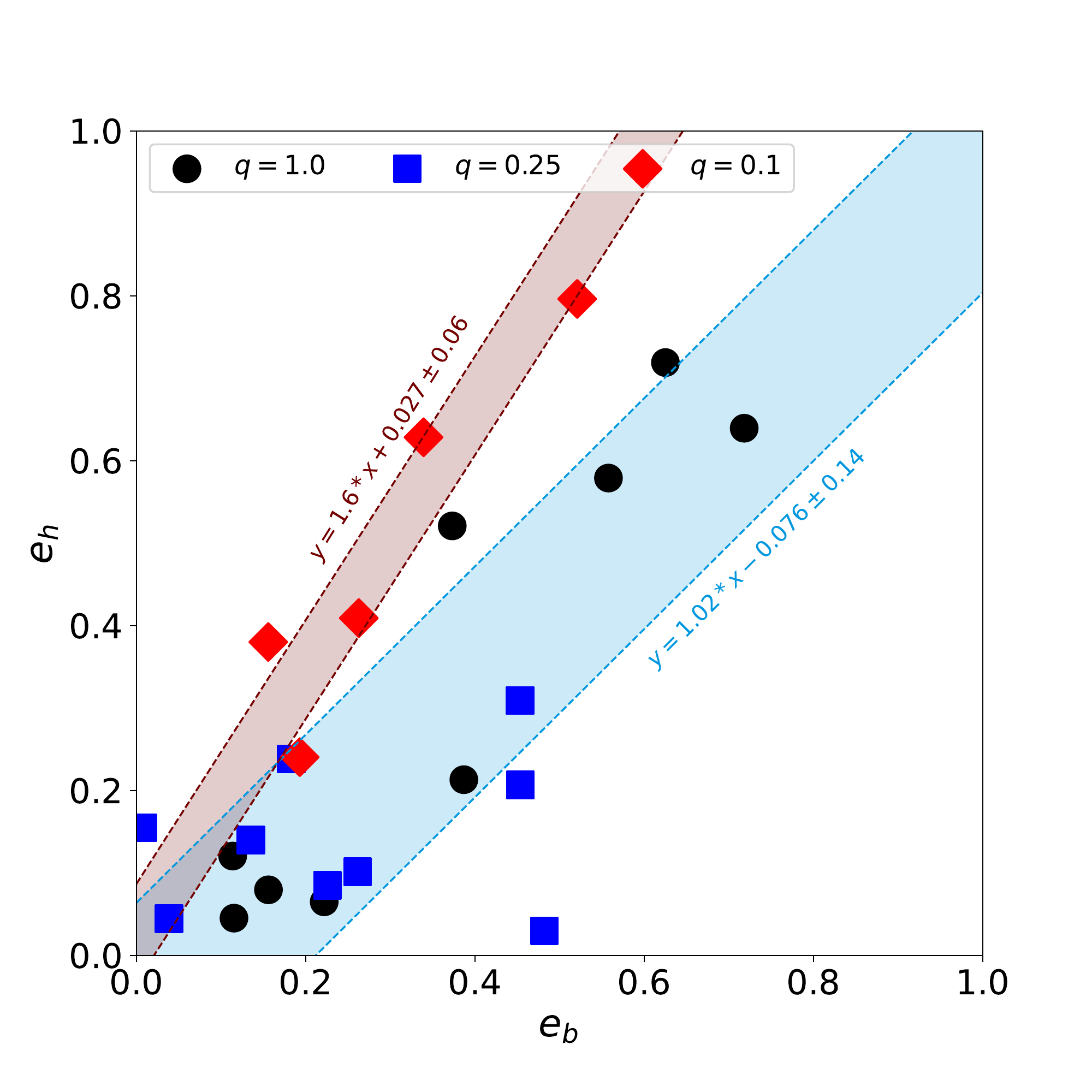}
    \caption{Correlation between the eccentricity measured at the hard-binary separation and that measured at the time of binding, as defined by Eq.\,\ref{eq:bound}, for models with different galaxy mass ratios. The shaded areas shows the $1\-\sigma$ confidence bands around the linear fits to the data for the $q=0.1$ models (red area) and $q=0.25, q=1$ models combined (blue area).}
    \label{fig:hard}
\end{figure}
The subsequent evolution does not alter the eccentricity significantly, and a clear correlation with the eccentricity at the hard-binary separation can be seen in Fig.\,\ref{fig:hard}, though with significant scatter. This implies that the main effect of the stochastic binding phase is to introduce a scatter in the dependence of the binary eccentricity at late times, after the hard-binary separation is reached, on the initial eccentricity of the galactic orbit. 

\subsection{Eccentricity evolution during the dynamical friction phase}
\label{sec: ecc_df}
As discussed in the previous section, the eccentricity of the orbit seems to constantly decrease during the dynamical friction phase until the formation of the bound binary. In order to investigate this phase in more detail, we adopt a semi-analytical approach. BY relying on simplified equations, the framework allows
to gain physical insight and explore a larger parameter space at a much lower computational cost with respect to $N$-body simulations.

We consider a setup in which the main host galaxy is modelled with a fixed analytical potential while the secondary MBH is represented by a massive perturber that sinks within the galaxy due to the action of a dissipative force mimicking dynamical friction. We note that this setup is inadequate for major mergers and we therefore restrict our comparison to the $q=0.1$ case, in which the merger does not significantly perturb the potential of the main galaxy. Specifically, we integrate the equations of motion of a massive particle affected by the conservative gravitational potential of a Dehnen mass distribution (with different values of $\gamma$) and by the dissipative dynamical friction  deceleration implemented as \citep{chandrasekhar1943}:  
%%%%%%%%%%%%%%%%%%
\begin{equation}\label{eq:DF_sph}
    \mathbf{a}_{\rm df} = -2\pi G^2 \ln(1+\Lambda^2) m_p \rho(r) \left({\rm erf}(X) - \dfrac{2 X{\rm e}^{-X^2}}{\sqrt{\pi}}\right) \dfrac{\mathbf{v}_{p}}{|\mathbf{v}_{p}|^3}.
\end{equation}
%%%%%%%%%%%%%%%%%%
In the above expression, $m_p$ and $\mathbf{v}_p$ denote the mass and velocity of the perturber, $\rho(r)$ is the local background density at the instantaneous radius $r$, $\Lambda = p_{\rm max}/p_{\rm min}$ is the ratio between the maximum and minimum impact parameter, while $X$ is a shortcut for $X=v_p/(\sqrt{2}\sigma(r))$ representing the ratio of the perturber velocity over the local velocity dispersion (at radius $r$ from the centre). For the maximum and minimum impact parameters we adopt the following expressions \citep{Just2011,Petts2016,Bonetti2021}:
%%%%%%%%%%%%%%%%%%
\begin{align}\label{eq:finesse}
    p_{\rm max} &= r/\delta, \nonumber \\
    p_{\rm min} &= \max\left(\dfrac{G m_p}{v_p^2+\sigma(r)^2}, D_p\right), \nonumber \\ 
    \delta &= - \dfrac{{\rm d} \ln \rho}{{\rm d} \ln r},
\end{align}
%%%%%%%%%%%%%%%%%%
$\delta$ being the logarithmic slope of the density profile, $r$ the radial coordinate and $D_p$ the physical radius of the infalling object, here assumed to be zero. The details of the orbital integrator can be found in \citet{Bonetti2020, Bonetti2021}.

We initialize the host system with total mass $M\approx 0.91$, scale radius $r_0=1$ and  $\gamma=0.5,1, 1.5$ (the same initial parameters as those of the main galaxy in the $q=0.1$ $N$-body runs), and the intruder point mass $m_p\approx4.55\times10^{-4}$ (equal to the mass of the secondary MBH in the $q=0.1$ $N$-body runs). We choose the initial orbit of the intruder so that its initial, non-Keplerian, orbital semi-major axis is fixed to $a=5$ and the non-Keplerian eccentricity is $e = 0.2, 0.4, 0.6, 0.8$; these  orbital parameters are again computed in such a way that the periapsis $r_p$ and apoapsis $r_a$ in absence of dynamical friction would satisfy the relations in Eq.~\ref{eq:peri_apo_params}.

\begin{figure}
    \centering
    \includegraphics[width=0.95\columnwidth]{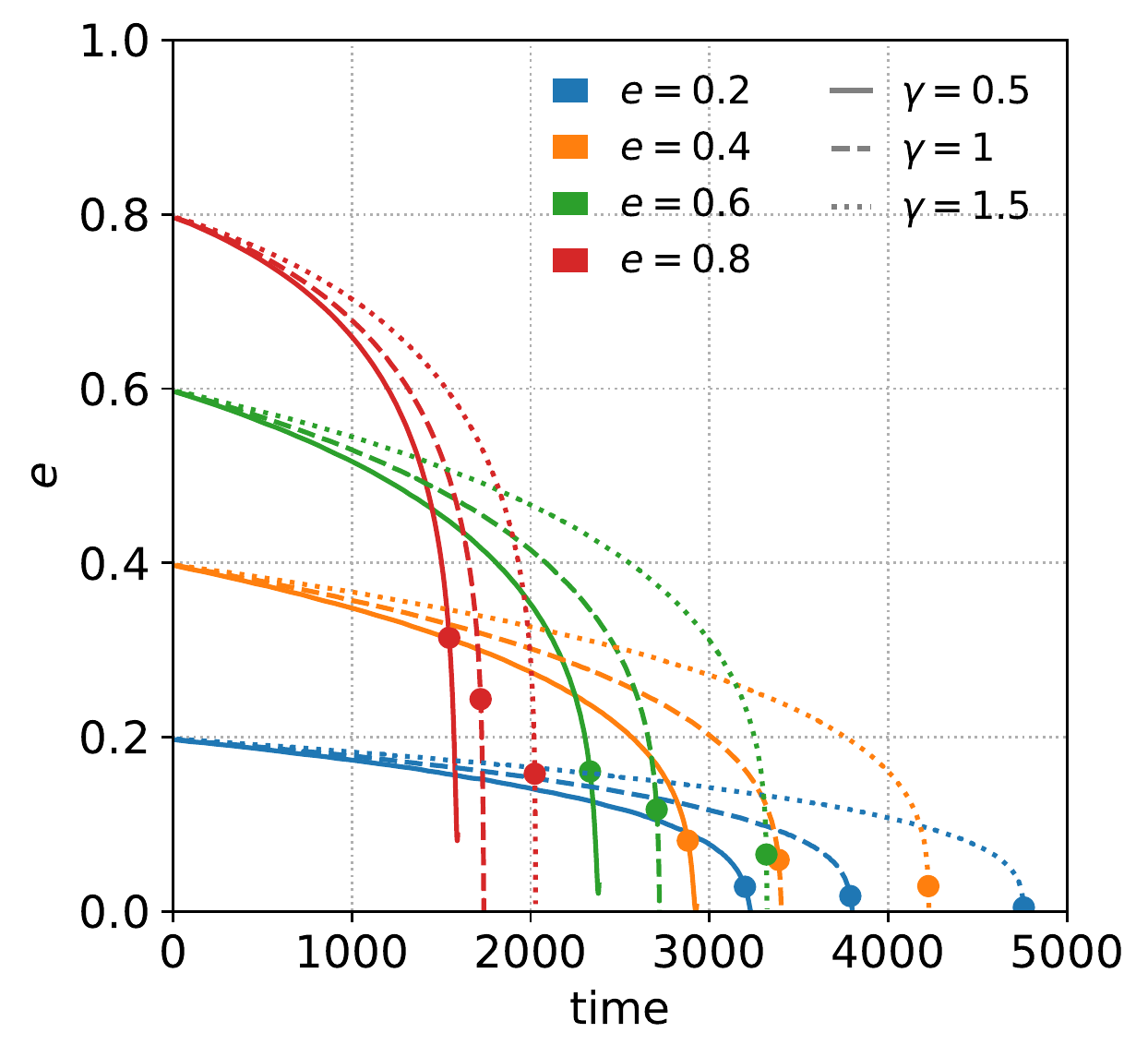}
    \includegraphics[width=0.95\columnwidth]{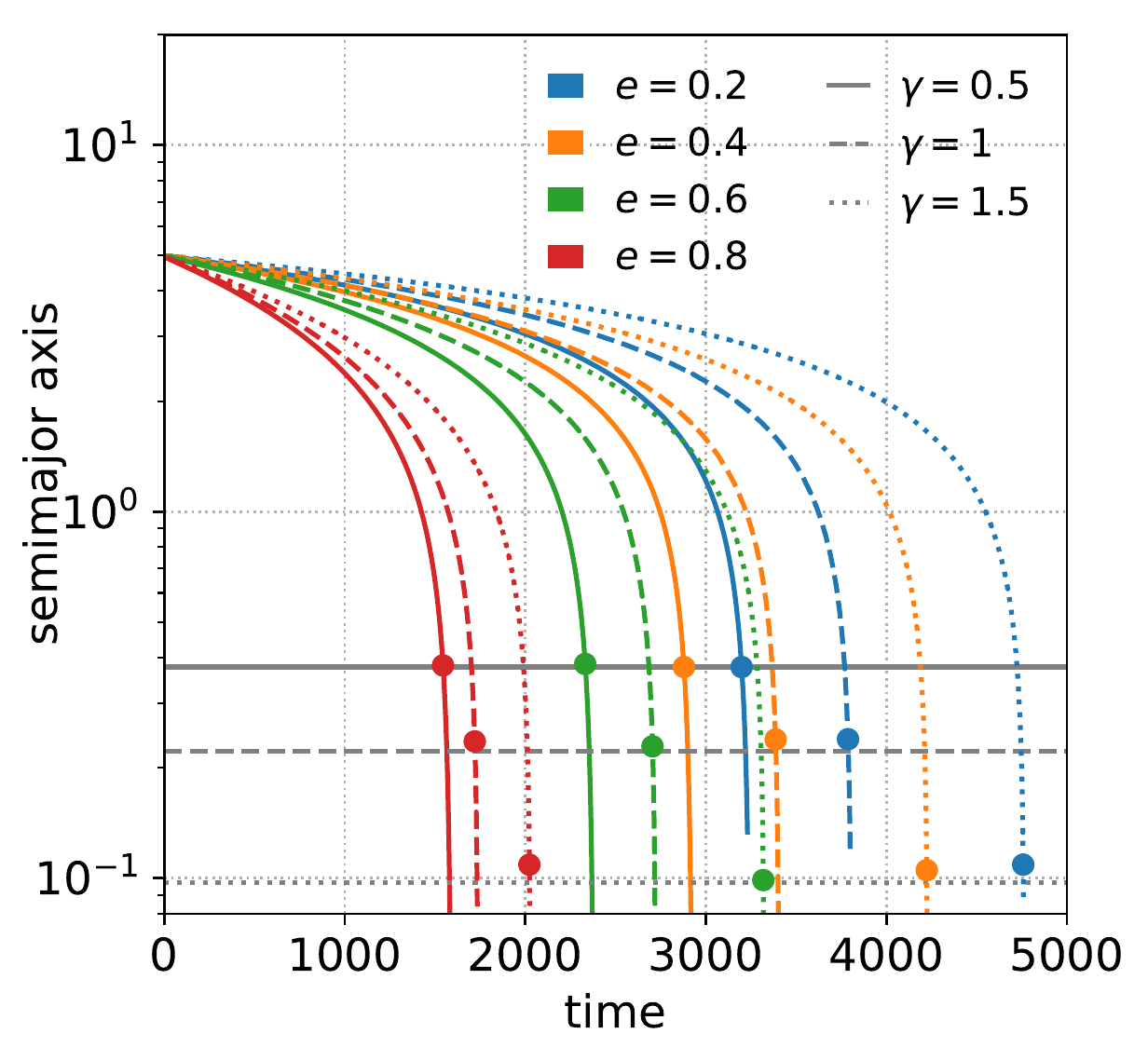}
    \caption{Evolution of the eccentricity (top) and semi-major axis (bottom) of a point mass object subject to dynamical friction within a Dehnen density profile. The orbits are initialized with different inner slopes for the host galaxy ($\gamma$) and different values for the initial orbital eccentricity. At any time, the eccentricity and semi-major axis are computed via Eq.~\ref{eq:peri_apo_params}. The dots on each line mark the time at which the massive object reaches a semi-major axis equal to twice the influence radius of a MBH with 10 times its mass (representative of the $q=0.1$ models in this study).}
    \label{fig:semianalytical}
\end{figure}

The results of the integration are shown in Fig.~\ref{fig:semianalytical}.
In the upper (lower) panel we show the time evolution of the eccentricity (semi-major axis) of the point mass object for several different initial eccentricities (from 0.2 to 0.8; see colour code) and three different slopes of the spherical density profile (0.5, 1.0, 1.5; see different line-styles). The non-Keplerian orbital parameters are computed as in Eq.~\ref{eq:peri_apo_params}. The evolution confirms the trend recovered in the full $N$-body simulations for which the eccentricity systematically decreases during the dynamical friction phase. Larger initial eccentricities yield, for a fixed $\gamma$, a faster orbital decay. This owes to the greater efficiency of dynamical friction at pericentre, so that more eccentric initial orbits can penetrate deeper into the stellar cusp and dissipate more compared to more circular ones; the fact that dynamical friction is more efficient at pericentre also implies that the object tends to circularize. For the same initial orbital eccentricity, inspirals take longer in steeper density profiles. This is due to the fact that in all considered cases the total mass is kept fixed at a constant value, therefore increasing the inner slope of the profile results in a slightly reduced density (and in turn dynamical friction efficiency) in the outer regions of the system ($r \gg r_0$), where most of the inspiral takes place. It is worth stressing that in this semi-analytical approach we can only account for the effect of dynamical friction, and we neglect the effect of mass loss of the intruder, which is instead treated as a point mass within the host potential. By comparing the eccentricity endpoints in figure \ref{fig:semianalytical} to those in table \ref{tab:data} for the $q=0.1$ cases, we can appreciate the consistency of the trends. Circulatization occurs in all cases. For a fixed initial eccentricity, $e_b$ is smaller for steeper cusps, whereas for a fixed cusp profile, initially more eccentric binaries result in a larger $e_b$. Note, however, that the semi-analytical model predicts longer DF timescales for steeper cusps, which is opposite to the trend seen in the simulations. This is because in the simulation the secondary MBH is not naked, but it is surrounded by an extended galaxy that is progressively stripped in the process. The steeper the cusp, the more concentrated the stellar profile and the less efficient the stripping. In reality, the MBH is surrounded by more stars in steeper cusps, leading to a decreasing DF timescale.

The circularization that we observe in the dynamical friction phase is in tension with the results by \citet[][]{vasiliev2021} who find an increase in the (orbital) eccentricity of two merging spherical systems, in the case of (i)  mass ratios not too far from unity, (ii) relatively shallow density profiles,
 and (iii) large initial eccentricity. This {\it radialization} is in contrast with what is typically found applying  \citeauthor{chandrasekhar1943}'s theory (Eq.~\ref{eq:DF_sph}). 
 They show that it is a complex process arising from a combination of different causes, most importantly the torque exerted by the stripped satellite debris on the satellite itself, and the recoil induced by the infalling satellite in the inner regions of the host (thus implying that \citeauthor{chandrasekhar1943}'s local treatment of this phenomenon may not be fully adequate). This probably implies that radialization is less likely to happen if the merging satellite is (i)  too compact  -- as it is less prone to stripping; (ii)  too diluted --  as the stripping at first periapsis would result, in our case, in a naked MBH whose low mass would not induce any recoil in the host and whose associated satellite debris would be too spread out to significantly induce a torque on the MBH; (iii) too light -- as both the recoil of the host and the self-induced torque would be weaker; (iv) initially on a relatively circular orbit --  as the effects that induce recoil are maximised the deeper the first pericentre is. This may help us understand why we never clearly see radialization in our $N$-body simulations and in fact we observe circularization in the dynamical friction phase, better matching the standard dynamical friction predictions: our satellites are more diluted compared to the ones in \citet{vasiliev2021}, and they initially have a larger pericentre passage, thus they may be more prone to induce a mainly local, instead of a global, response in the system. In addition, part of the difference might be due to the single versus double power-law density profiles adopted in their and our study, respectively.
 
 \subsection{Eccentricity evolution in the binding phase}
 In the stage of binary formation and in the immediate aftermath, the binary experiences a phase of rapid shrinking mainly due to the ejection of stars bound to the two MBHs. This phase brings the binary from a separation $a_b$ to $a_h$ and results in the scouring of a stellar core. The eccentricity evolution in the core-scouring phase has been investigated in \cite{Sesana2010} with an hybrid approach combining 3-body scattering experiments and a semi-analytical framework for the evolution of the stellar distribution. Their formalism is strictly applicable only to unequal mass binaries ($q\lesssim 0.1$), but it has been extended to higher mass ratios. They find that the scouring phase does not significantly alter the eccentricity if $q$ is close to unity, whereas for $q\approx 0.1$ an increase in the eccentricity is expected, with shallower cusps leading to larger growth. 
 
 Values of eccentricity at binding $e_b$ and the hard-binary separation $e_h$ found in our simulations are listed in Tab. \ref{tab:data} and shown in Fig. \ref{fig:hard}. The results qualitatively corroborate the predictions of the analytical models. Mergers with $q=0.25$ and $q=1$ roughly follow a linear trend $e_b \sim e_h$, albeit with significant scatter and one obvious outlier. We performed a Deming regression and obtained a best linear fit $e_h=\alpha e_b + \beta$ with $\alpha=1.02$ and $\beta=-0.076$, and intrinsic dispersion of the relation of $\epsilon=0.103$, which is visualised in figure \ref{fig:hard}. We immediately notice that the relation is consistent with $e_b=e_h$ `within one sigma', although with a small bias towards circularization, $e_h<e_b$. Furthermore, the five simulations with $q=0.1$ lie outside the upper $1\sigma$ dispersion region covered by the relation. If those values were produced by the above relation, each of them would have a probability of $\approx 0.16$ to be above the 1$\sigma$ region. The probability to get five points over this line is $\approx 0.16^5\approx 10^{-4}$. Therefore the distribution of the $q=0.1$ points in figure \ref{fig:hard} is a $4\sigma$ outlier of five independent draws from the fitting relation.
 The $q=0.1$ points, instead, closely follow a linear relation with $\alpha=1.6$ and $\beta=0.027$, with a smaller $\epsilon=0.038$ intrinsic dispersion. Although based on a handful of points, these relations can be used to statistically predict the eccentricity of a binary at the hardening separation.
 
 A prediction of the hybrid models based on 3-body scatterings is that, for binaries with small $q$, the eccentricity growth in the core scouring process should depend on the initial slope $\gamma$ of the cusp, with shallower cusps leading to larger growth. Again, this trend is found in our simulations with $q=0.1$. For $\gamma=0.5$ the only simulation reaching $a_h$ experiences a variation $\delta e=0.28$ in this phase. The average growth for $\gamma=1$ is $\delta{e}=0.25$, whereas for $\gamma=1.5$ is only $\delta{e}=0.1$.  Although a direct comparison is hard, results shown in figure 3 and 4 of \cite{Sesana2010} for $q=1/9$ display comparable growth, $\delta{e}\approx 0.1-0.4$. Note that in the same figures the eccentricity evolution of equal mass binaries in this phase is consistent with zero regardless of the other properties of the system.
 
\subsection{Eccentricity evolution during the hardening phase}

\begin{figure}
    \includegraphics[width=1.\columnwidth]{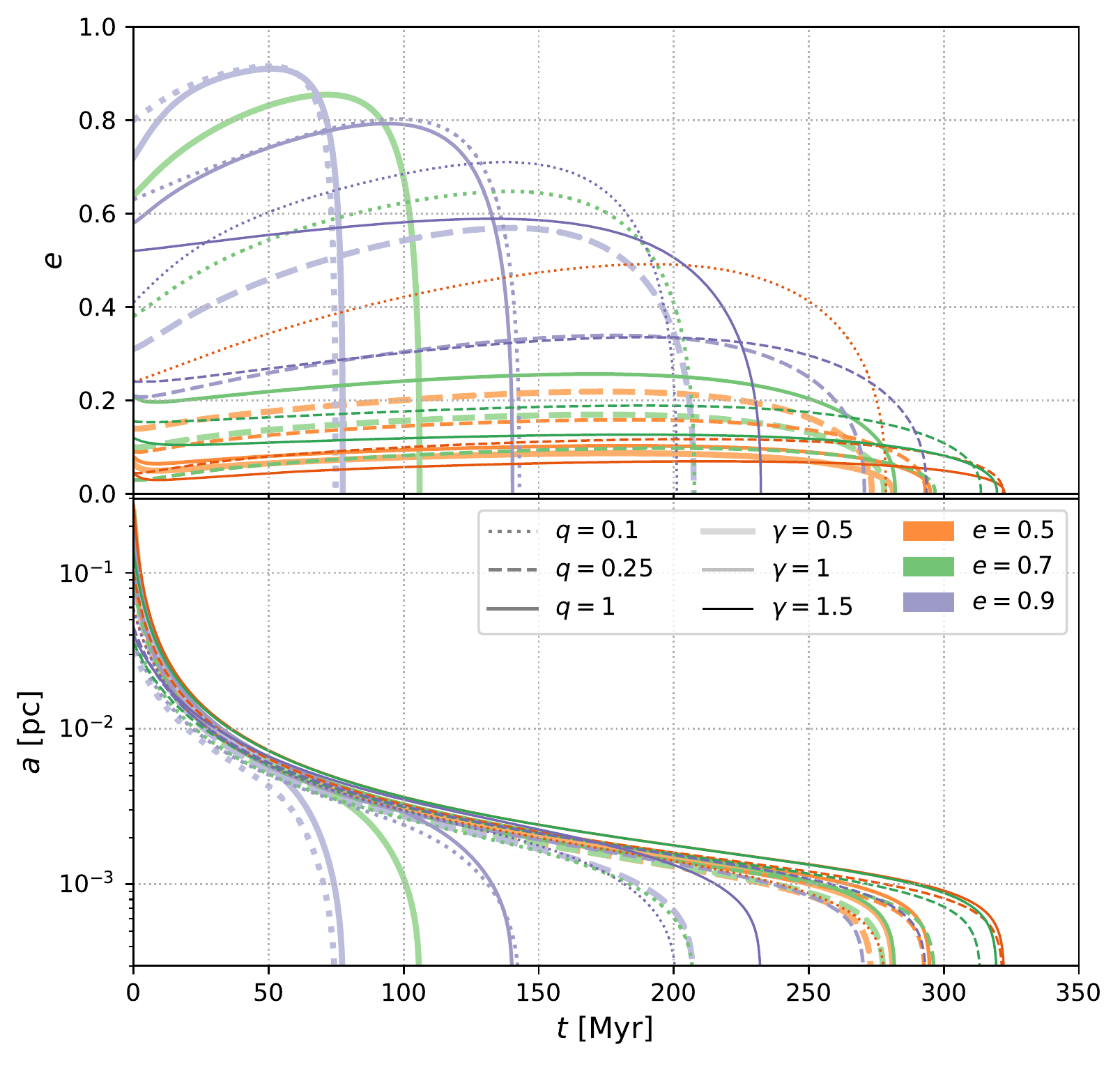}
        \caption{The image shows the evolution of the binary eccentricity (top) and semi-major axis (bottom) as a function of time as obtained from solving Eqs.~(\ref{eq:hardgwa}, \ref{eq:hardgwe}), using the initial conditions for $a_h$ and $e_h$ as in Tab.~\ref{tab:data}. Note that here $t=0$ corresponds to $t=t_h$, and the tracks have been obtained assuming the scaling in Tab.~\ref{tab:units} with $M_b=4\times10^6\msun$. Dotted, dashed and solid lines respectively refer to runs with $q=0.1, 0.25, 1$; wider and lighter lines to thicker and darker ones respectively refer to $\gamma = 0.5, 1, 1.5$; red, green and violet lines refer to runs with initial eccentricities equal to $0.5, 0.7, 0.9$.}
    \label{fig:hardgw}
\end{figure}
We stop our $N$-body simulations shortly after the binary has reached the hard binary separation,  $a_h$. From this point onward we can integrate the evolution of each binary in a semi-analytical fashion. We assume that the evolution is driven by both stellar interactions and GW emission, with the latter becoming dominant at late times. The rate of change of the orbital elements can therefore be written as
\begin{align}\label{eq:hardgwa}
    \frac{da}{dt} &= \left. \frac{da}{dt} \right|_\star + \left.\frac{da}{dt}\right|_{\rm GW}\\
    \frac{de}{dt} &= \left.\frac{de}{dt}\right|_\star + \left.\frac{de}{dt}\right|_{\rm GW}\label{eq:hardgwe}\,.
\end{align}
The evolution mediated by stellar interactions, denoted with a star symbol in the equations above, is well described by \citep[][]{quinlan1996}
\begin{align}
\label{eq:eqh}
    \left. \frac{da}{dt} \right|_\star  = -a^2\frac{HG\rho}{\sigma}\\
    \left. \frac{de}{dt} \right|_\star = a \frac{HKG\rho}{\sigma}
\end{align}
where $\rho$, $\sigma$ are the stellar density and velocity dispersion at the binary influence radius, while $H$, $K$ are coefficients that depend on the binary mass ratio, separation and eccentricity, and can be derived by means of scattering experiments; here we use the fits obtained by \citet[][]{sesana2006} under the assumption of an isotropic stellar background. 
The GW-driven evolution of the orbital elements can be modelled via \citep[][]{peters1964}
\begin{align}
\label{eq:eqgw}
    \left. \frac{da}{dt} \right|_{\rm GW}  &= -\frac{64 G^3}{5c^5}\frac{M_1M_2}{a^3(1-e^2)^{7/2}}\left(1+\frac{73}{24}e^2+\frac{37}{96}e^4\right) \\
    \left. \frac{de}{dt} \right|_{\rm GW} &=  -\frac{304 G^3}{15c^5}\frac{M_1M_2}{a^4(1-e^2)^{5/2}}\left(e+\frac{121}{304}e^3\right);
\end{align}
here $c$ is the speed of light in vacuum.
In order to evolve the system past the hard binary separation, we need to set a value for $\rho/\sigma$; for each run, we evaluate this ratio at a radius enclosing a mass in stars equal to twice the binary mass at $t=t_h$. Empirically, we find that this agrees within $\approx 20$ per cent with $\tfrac{1}{2}\rho/\sigma$ evaluated at a radius containing $0.5\%$ of the mass of a \citeauthor{Dehnen1993} model with total mass equal to $M_{\rm tot}=1$ and  $r_0$, $\gamma$ equal to the ones chosen for the initialization of each run. In order to evolve the set of differential equations given above (Eq.\,\ref{eq:eqh} and \ref{eq:eqgw}) we also need to select a set of physical units. We first consider the scaling shown in Tab.~\ref{tab:units} for a $4\times10^6 \msun$ central binary. The obtained tracks, initialised at $a_h$ and $e_h$ from Tab.~\ref{tab:data}, are shown in Figure \ref{fig:hardgw}. This scaling assumes that the density at the edge of the influence sphere is $\approx 5000 - 8000 \msun \pc^{-3}$, implying a relatively efficient hardening that leads to coalescence in less than 350 Myr. Rescaling to the case of a $10^8 \msun$ binary, the time elapsed from $t_h$ to coalescence is approximately $100-500$ Myr, and the maximum eccentricity attained in each run is slightly smaller than the ones in Fig.~\ref{fig:hardgw}. In this case, the density at the edge of the influence sphere is around $500-800\msun \pc^{-3}$.

\begin{table}
\begin{center}
\caption{Hardening phase: identifier, maximum eccentricity achieved during the hardening phase, physical time elapsed from the start of the simulation to the hard-binary separation, physical time elapsed from the hard-binary separation to coalescence, total physical time to coalescence, computed for two adopted scalings: (A) a binary of mass $4\times10^6\msun$ and (B) $10^8\msun$.}
\label{tab:hard}
\begin{tabular}{lcccc} 
\hline
Name & $e_{\rm max}$ & $T_{\rm hard}$ (yr) & $T_{\rm merg}$ (yr) & $T_{\rm tot}$ (yr)\\
\hline
(A)   G05E05Q1  & 0.09  & $2.81\times10^8$ & $2.63\times10^7$ & $3.08\times10^8$\\
(B)   G05E05Q1  & 0.07  & $4.15\times10^8$ & $7.86\times10^7$ & $4.93\times10^8$\\
(A) G05E05Q025  & 0.22  & $2.73\times10^8$ & $7.17\times10^7$ & $3.45\times10^8$\\
(B) G05E05Q025  & 0.19  & $4.05\times10^8$ & $2.14\times10^8$ & $6.19\times10^8$\\
(A)   G05E07Q1  & 0.85  & $1.06\times10^8$ & $1.82\times10^7$ & $1.24\times10^8$\\
(B)   G05E07Q1  & 0.80  & $1.92\times10^8$ & $5.45\times10^7$ & $2.47\times10^8$\\
(A) G05E07Q025  & 0.17  & $2.78\times10^8$ & $5.85\times10^7$ & $3.36\times10^8$\\
(B) G05E07Q025  & 0.15  & $4.10\times10^8$ & $1.75\times10^8$ & $5.85\times10^8$\\
(A)   G05E09Q1  & 0.91  & $7.75\times10^7$ & $1.13\times10^7$ & $8.88\times10^7$\\
(B)   G05E09Q1  & 0.88  & $1.40\times10^8$ & $3.37\times10^7$ & $1.73\times10^8$\\
(A) G05E09Q025  & 0.57  & $2.07\times10^8$ & $3.09\times10^7$ & $2.38\times10^8$\\
(B) G05E09Q025  & 0.48  & $3.35\times10^8$ & $9.24\times10^7$ & $4.27\times10^8$\\
(A)  G05E09Q01  & 0.92  & $7.48\times10^7$ & $1.66\times10^8$ & $2.41\times10^8$\\
(B)  G05E09Q01  & 0.88  & $1.33\times10^8$ & $4.95\times10^8$ & $6.29\times10^8$\\
(A)    G1E05Q1  & 0.10  & $2.95\times10^8$ & $5.88\times10^7$ & $3.54\times10^8$\\
(B)    G1E05Q1  & 0.09  & $4.35\times10^8$ & $1.76\times10^8$ & $6.11\times10^8$\\
(A)  G1E05Q025  & 0.16  & $2.93\times10^8$ & $1.22\times10^8$ & $4.15\times10^8$\\
(B)  G1E05Q025  & 0.14  & $4.32\times10^8$ & $3.65\times10^8$ & $7.97\times10^8$\\
(A)    G1E07Q1  & 0.26  & $2.82\times10^8$ & $3.91\times10^7$ & $3.21\times10^8$\\
(B)    G1E07Q1  & 0.23  & $4.20\times10^8$ & $1.17\times10^8$ & $5.37\times10^8$\\
(A)  G1E07Q025  & 0.10  & $2.97\times10^8$ & $9.51\times10^7$ & $3.92\times10^8$\\
(B)  G1E07Q025  & 0.08  & $4.36\times10^8$ & $2.84\times10^8$ & $7.20\times10^8$\\
(A)   G1E07Q01  & 0.65  & $2.08\times10^8$ & $6.26\times10^8$ & $8.33\times10^8$\\
(B)   G1E07Q01  & 0.57  & $3.39\times10^8$ & $1.87\times10^9$ & $2.21\times10^9$\\
(A)    G1E09Q1  & 0.79  & $1.40\times10^8$ & $2.08\times10^7$ & $1.61\times10^8$\\
(B)    G1E09Q1  & 0.73  & $2.45\times10^8$ & $6.22\times10^7$ & $3.08\times10^8$\\
(A)  G1E09Q025  & 0.34  & $2.71\times10^8$ & $4.95\times10^7$ & $3.20\times10^8$\\
(B)  G1E09Q025  & 0.28  & $4.10\times10^8$ & $1.48\times10^8$ & $5.58\times10^8$\\
(A)   G1E09Q01  & 0.80  & $1.43\times10^8$ & $2.57\times10^8$ & $4.00\times10^8$\\
(B)   G1E09Q01  & 0.73  & $2.52\times10^8$ & $7.67\times10^8$ & $1.02\times10^9$\\
(A)   G15E05Q1  & 0.07  & $3.22\times10^8$ & $2.22\times10^8$ & $5.45\times10^8$\\
(B)   G15E05Q1  & 0.06  & $4.75\times10^8$ & $6.64\times10^8$ & $1.14\times10^9$\\
(A) G15E05Q025  & 0.12  & $3.22\times10^8$ & $3.59\times10^8$ & $6.81\times10^8$\\
(B) G15E05Q025  & 0.10  & $4.74\times10^8$ & $1.07\times10^9$ & $1.55\times10^9$\\
(A)  G15E05Q01  & 0.49  & $2.79\times10^8$ & $1.15\times10^9$ & $1.43\times10^9$\\
(B)  G15E05Q01  & 0.40  & $4.42\times10^8$ & $3.45\times10^9$ & $3.89\times10^9$\\
(A)   G15E07Q1  & 0.13  & $3.20\times10^8$ & $1.43\times10^8$ & $4.63\times10^8$\\
(B)   G15E07Q1  & 0.12  & $4.71\times10^8$ & $4.28\times10^8$ & $8.99\times10^8$\\
(A) G15E07Q025  & 0.19  & $3.14\times10^8$ & $2.96\times10^8$ & $6.10\times10^8$\\
(B) G15E07Q025  & 0.17  & $4.56\times10^8$ & $8.84\times10^8$ & $1.34\times10^9$\\
(A)   G15E09Q1  & 0.59  & $2.32\times10^8$ & $9.72\times10^7$ & $3.29\times10^8$\\
(B)   G15E09Q1  & 0.55  & $3.52\times10^8$ & $2.90\times10^8$ & $6.43\times10^8$\\
(A) G15E09Q025  & 0.34  & $2.94\times10^8$ & $1.51\times10^8$ & $4.45\times10^8$\\
(B) G15E09Q025  & 0.28  & $4.39\times10^8$ & $4.52\times10^8$ & $8.91\times10^8$\\
(A)  G15E09Q01  & 0.71  & $2.01\times10^8$ & $7.32\times10^8$ & $9.33\times10^8$\\
(B)  G15E09Q01  & 0.63  & $3.40\times10^8$ & $2.19\times10^9$ & $2.53\times10^9$\\
\hline
\end{tabular} 
\end{center}
\end{table}
We list the time spent, in physical units, from the beginning of the simulation to the hard-binary separation $T_{\rm hard}$, from the hard-binary separation to coalescence $ T_{\rm merg}$ and the total time to coalescence $T_{\rm tot}$ in Table \ref{tab:hard}, for both adopted scalings, together with the maximum eccentricity achieved by the binary in the hardening phase. We note that for most cases $T_{\rm hard} > T_{\rm merg}$, the only notable exceptions being the $q=0.1$ models, for which $T_{\rm merg} \sim 5-10 \,T_{\rm hard}$, due to the inefficiency of dynamical friction for light secondary MBHs.
Total elapsed times from the onset of the galactic merger to coalescence of the MBHs span the range of a few hundred Myr to a a few Gyr, depending on the parameters. The shortest time of $\lesssim 100$ Myr is recorded for model G05E09Q1 which is an equal mass merger with a shallow density profile and a large initial orbital eccentricity.

It is very important to stress that the results obtained in this Section are very sensitive to the scaling adopted. If we were to set a length unit so that the central density were very low ($\lesssim 10 \msun \pc^{-3}$) the binary would take more than a Hubble time to coalesce;  the opposite is true if the central density were very high. This means that the total time to coalescence depends crucially on the adopted scaling and comparisons with other works must take this into account.

We note that the eccentricity is predicted to grow in the hardening phase due to encounters with passing stars \citep{sesana2006,Sesana2010}, unless the mass ratio is very small \citep{Bonetti2020a}. The semi-analytical implementation reproduces this behaviour, as can be appreciated from Fig.~\ref{fig:hardgw}.

\subsection{Stochasticity and convergence}
\label{sec:noise}
Stochasticity in the eccentricity of BHBs has been reported in \cite{nasim2020} owing to the intrinsically chaotic nature of the interactions between the BHs and the stars during the merger and the hardening phase. This results in a significant scatter in the extrapolated merger timescales of BHBs formed in gas free major mergers that scales with resolution as a Poissonian process. 

\begin{figure}
    \centering
    \includegraphics[width=1.0\columnwidth]{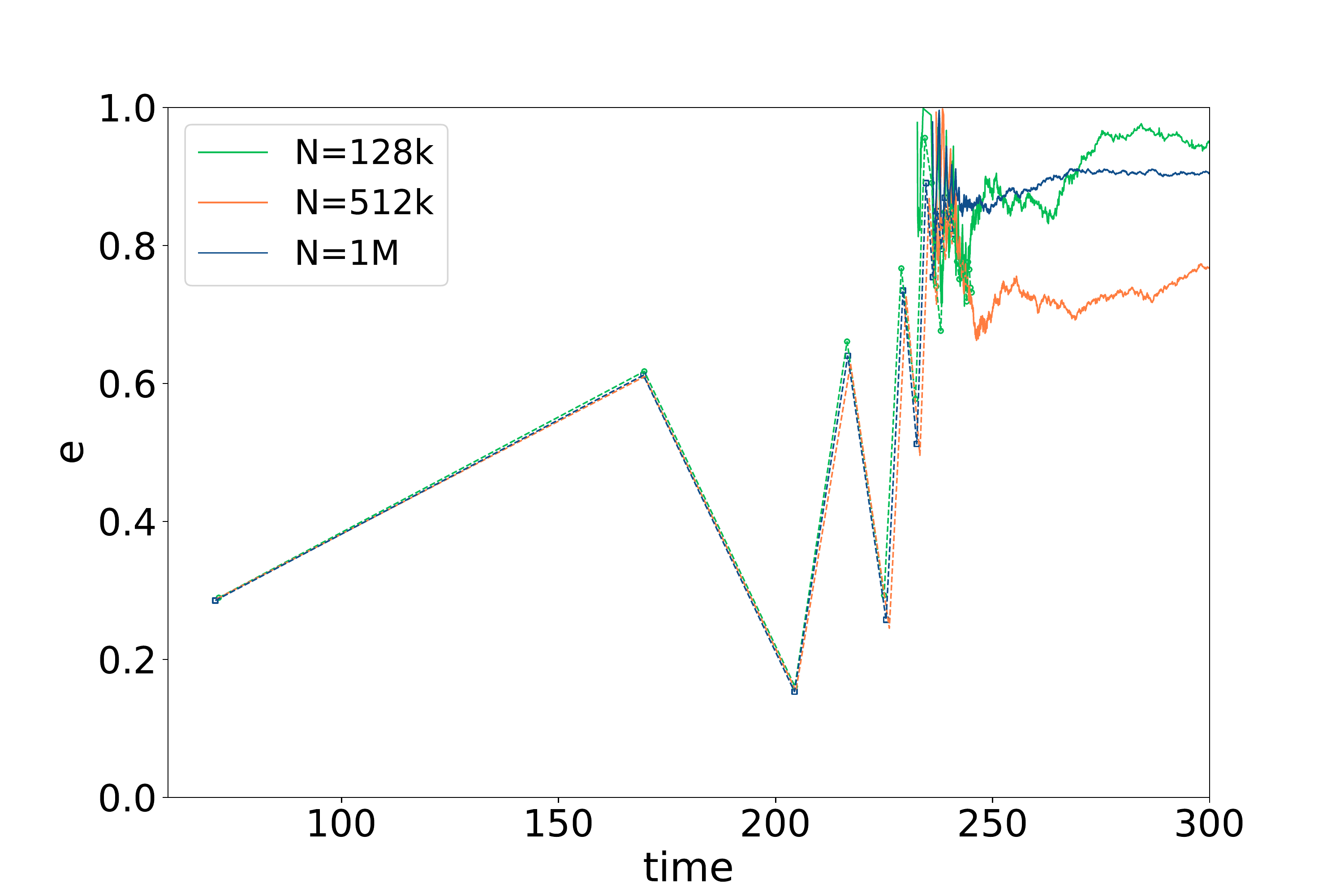}
        \caption{Eccentricity evolution as a function of time for model G05E07Q1 run at different particle numbers. The dashed lines show the orbital eccentricity computed from the pericentre and apocentre of the BHB while the solid lines show the Keplerian 2-body eccentricity. The observed scatter is smaller than that reported in \citet{nasim2020}, consistent with the expectation that perturbations and stochasticity are largest for nearly radial galaxy orbits.}
    \label{fig:conv}
\end{figure}
In order to assess the effects of stochasticity on our determination of the bound eccentricity, we repeat run G05E07Q1 with 5 different random realisations and compare the resulting eccentricity of the BHB at the first pericentre passage $e_p$, at the time of binding $e_b$ and at the time the hard binary separation is reached $e_h$. We find $e_p = 0.608 \pm 0.001$, $e_b = 0.795 \pm 0.09$ and $e_h = 0.76 \pm 0.07$. As expected, the eccentricity at the first pericentre passage shows a very small scatter, while a more significant, though modest, scatter is observed for the eccentricity measured at later times. Because the scatter owes to the stochasticity of stellar encounters with the BHB, it scales with particle number. We repeat run G05E07Q1 at higher (N=1M) and lower (N=128k) particle number and perform the same analysis on the eccentricity. While we can't properly compare with the scatter reported in \citet{nasim2020} without performing multiple runs at larger $N$,  Fig.\ref{fig:conv} shows a smaller deviation in our models. 
We attribute this result to the smaller initial orbital eccentricity of the merger ($e=0.7$ compared to $e=0.9$ in \citet{nasim2020}), as nearly radial orbits are more susceptible to perturbations.

\subsection{Flips}
All binaries in our simulation suite begin as co-rotating with respect to the stellar system. However, as described in \citet{Bortolas2018} and later studied in \citet{nasim2021}, the most eccentric binaries flip their angular momentum and reverse their rotation. We observe this effect in models G05E09Q1 and G1E09Q1 which have shallow or mildly cuspy initial profiles and large initial orbital eccentricities. 
\begin{figure*}
    \centering
    \includegraphics[width=1.0\columnwidth]{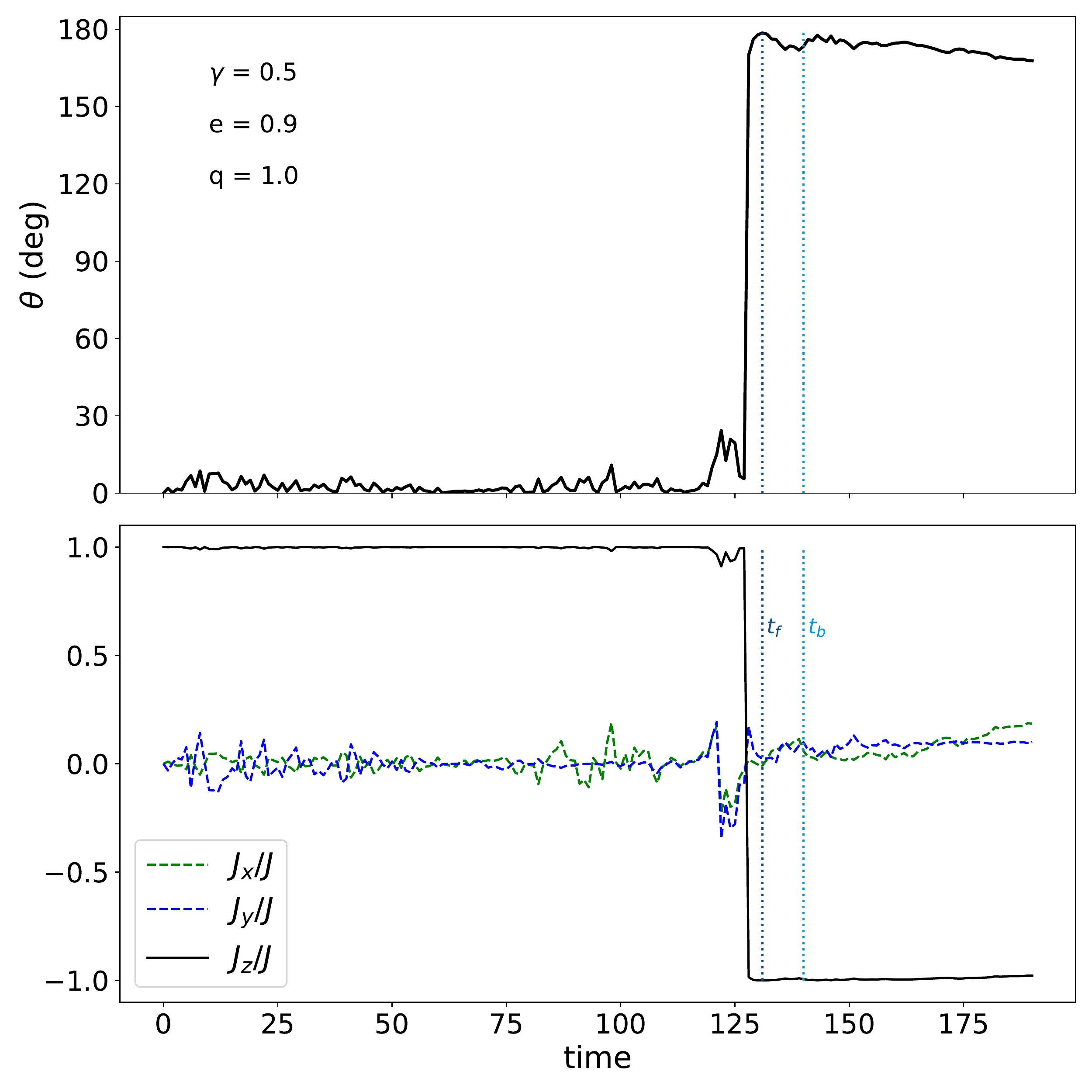}
    \includegraphics[width=1.0\columnwidth]{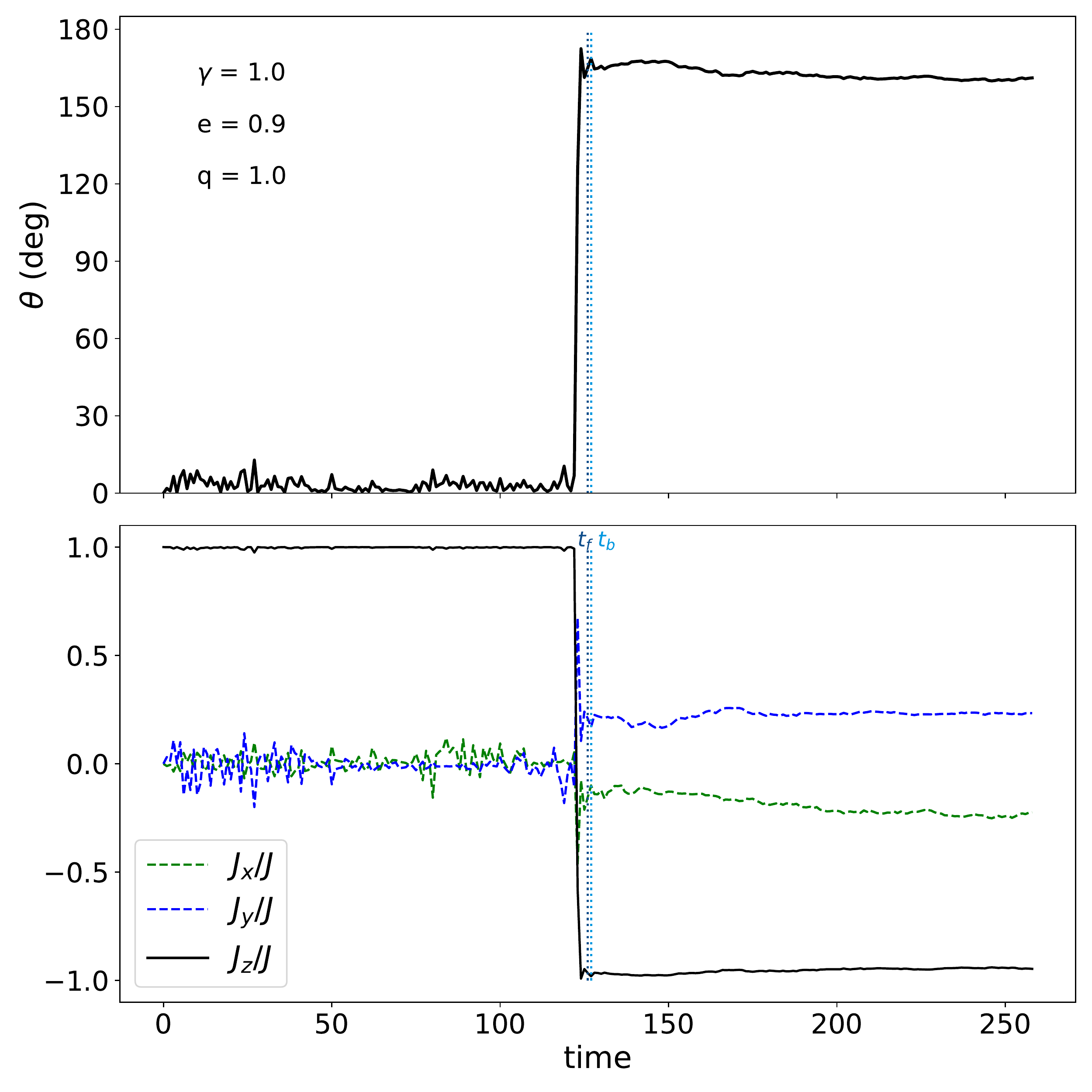}
        \caption{(Top) Evolution of the angle between the angular momentum vector of the BHB and the total angular momentum vector of the stars, for two models with high initial eccentricity ($e=0.9$) and different density slopes ($\gamma=0.5$ (left) and $\gamma=1$ (right)). (Bottom) Evolution of the components of the angular momentum vector of the BHB $J_x, J_y, J_z$, normalised to the magnitude of the vector $J$. The BHBs are initially co-rotating with the stellar distribution ($\theta \sim 0$) but show a fast flip to counter-rotation.}
    \label{fig:flip}
\end{figure*}
Fig.\ref{fig:flip} shows the evolution of the angle between the angular momentum vector of the BHB and the total angular momentum vector of the stars for the two models, as well as the evolution of the components of the angular momentum vector of the BHB. 
\citet{nasim2020} find that the flips occur around the time of binary formation, though they do not provide an explanation. They attribute the flips to torques from the triaxial merger remnant, which are unrelated to the time of binary formation. 
Our careful analysis of the binding process shows that, instead, flips occur earlier in the evolution and seem to be associated with the end of the merger rather than the binding of the BHs. In time, they are closer to the time $t_f$ which marks the end of the dynamical friction phase and is known to mark the end of the merging process, though they occur even earlier than $t_f$. 
We only observe flips in the models with large orbital eccentricity 
($e=0.9$) and a shallow central density profile ($\gamma=0.5, 1$). \citet{Bortolas2018} show that these models are characterised by a mildly oblate or triaxial shape in the innermost regions (within $25\%$ of the enclosed stellar mass), followed by an extended phase of oscillations in the axis ratios until an oblate shape is reached. It seems therefore plausible to attribute the flips to torques from the aspherical merger remnant. Models with high concentration ($\gamma=1.5$), on the other hand, are more compact and evolve quickly towards an oblate shape. We compute the axis ratios $b/a$ and $c/a$  following the procedure in \citet{Bortolas2018}, where $a>b>c$ are the axes of the ellipsoid used to approximate the stellar distribution. The shortest axis $c$ is always perpendicular to the plane of the merger. 
\begin{figure}
    \centering
    \includegraphics[width=1.0\columnwidth]{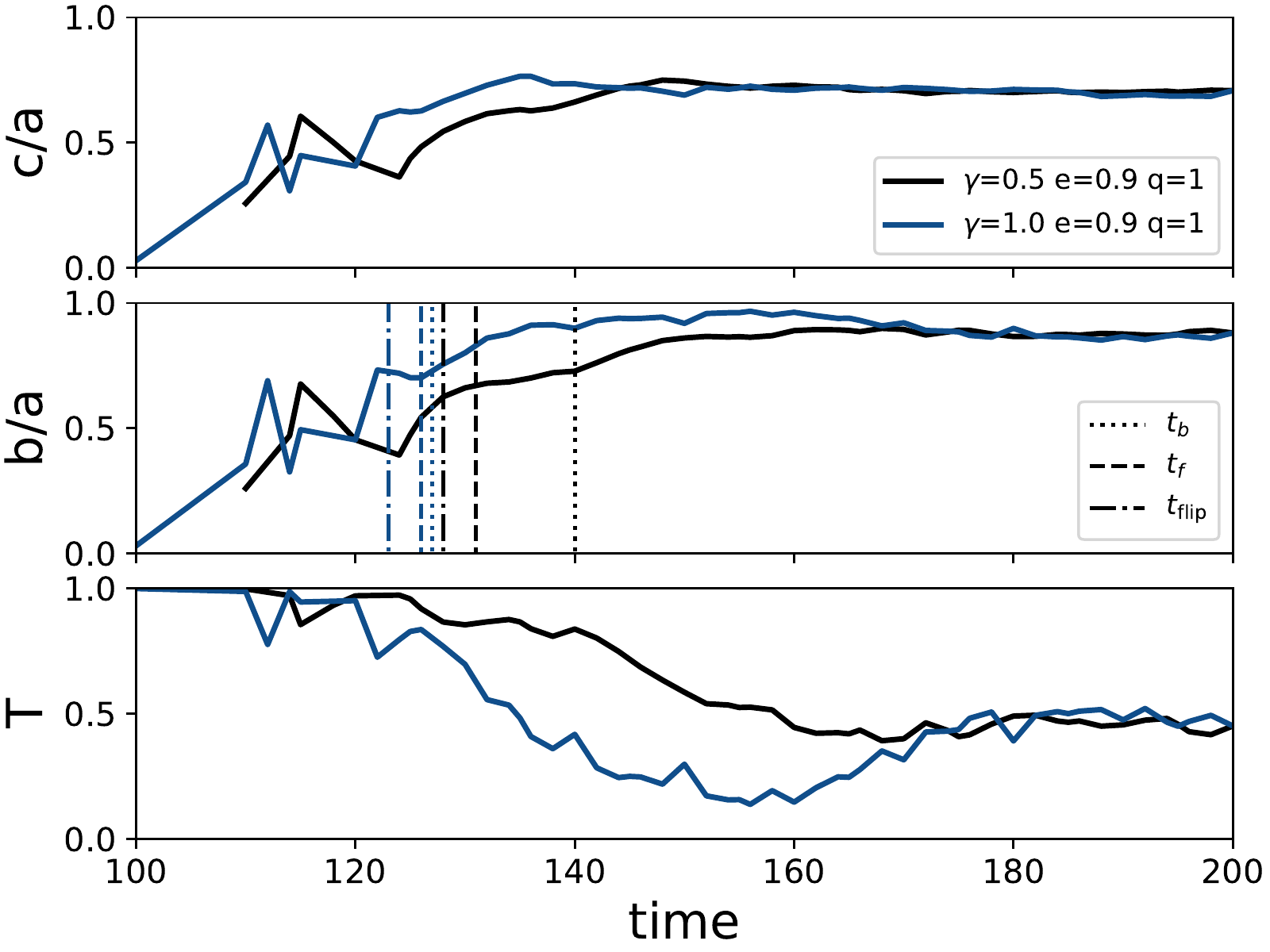}
            \caption{Axis ratios and triaxiality parameter as a function of time for models G05E09Q1 and G1E09Q1. These are computed using particles within a sphere enclosing a fraction of $\%$ of the total stellar mass, depending on the time. We find approximate agreement between the time the merger remnant reaches significant triaxiality and the time of the flips. }
    \label{fig:triax}
\end{figure}
The triaxiality parameter is then given by
\begin{equation}
    T = \frac{a^2-b^2}{a^2-c^2}
\end{equation}
with $0\leq T < 0.5$ representing an oblate spheroid, $T=0.5$ a maximally triaxial system and $0.5 < T \leq 1$ a prolate spheroid.
The shape of the remnant changes with radius, or equivalently mass enclosed within the BHB. We are interested in how triaxiality in the vicinity of the BHB is established during the merger and whether flips occur once a significant departure from spherical symmetry is in place. We compute the axis ratios and the triaxiality parameter as a function of time, starting at an early enough time that the merger is still underway and the BHs not yet bound. As shown in Fig.\,\ref{fig:triax}, the time of the recorded flips matches the phase of the merger where significant triaxiality is present in the central regions (between $25\%$ and $35\%$ of the total enclosed stellar mass), supporting the theoretical interpretation presented in \citet{nasim2021}. Models with a steeper profile and/or lower initial orbital eccentricity show axisymmetric rather than triaxial shapes, and no flips are observed. Reversing of the BHB's orbital plane is therefore an occurrence limited to a specific set of initial conditions.
Following a flip, the binary is counter-rotating with respect to the stellar distribution and therefore subject to torques that tend to realign its orbital angular momentum with the total angular momentum of the galaxy \citep{sesana2011, GDS2012}.

\section{Comparison with previous work}
The literature on the evolution of BHBs in stellar dominated environments is rich, although the main focus has often been the behaviour of the binary semi-major axis. It is, however, interesting to compare our findings with the relevant literature. The eccentricity at binding may appear lower than found in previous works. \citet{preto2011} find eccentricities as large as 0.95 in their merger simulations of spherical nuclei. However, this is only when considering MBHs lighter than assumed in this study ($M/M = 0.0025$) and near radial initial orbits. For $M/M = 0.005$ they find much more modest eccentricities at formation ($e \sim 0.7$). 
\citet{khan2012} find eccentricities in the range $0.5-0.98$ depending on the mass ratio and initial density slope. The discrepancy is due partly to our definition of bound eccentricity which removes the initial chaotic phase of binding and partly to differences in initial conditions. Even when the initial orbital eccentricity of the merger is comparable, differences may arise due to the location in the orbit where the BHs are placed, which leads to different pericentre distances. An additional source of disparity may be the choice of the softening parameter; this suppresses collisionality in the system and therefore the importance of close encounters. This in turn affects the angular momentum transfer between the stars and the BHB that drives the eccentricity evolution. We employ a much smaller softening for MBH-star interactions than other works, which may result in a lower eccentricity at formation. 

The eccentricity evolution in the dynamical friction phase has been studied by \citet{DA2017} for binaries with very small mass ratios ($ 10^{-4} \lesssim q \lesssim 10^{-1}$) evolving in stellar cusps of varying slope. The eccentricity is found to increase for any cusp shallower than $\gamma < 2$ at early times, when the binary is bound but not yet hard, and this is confirmed analytically assuming a description of dynamical friction which includes the contribution of stars faster than the infalling object. This might explain the short phase of increasing eccentricity that we observe in some of the $q=0.1$ models, our lightest infaller models, just after binding. They also predict a dependence of the eccentricity growth on the binary mass ratio, arguing that lighter intruders are able to reach a larger eccentricity by the time $\ah$ is reached. This might explain our observed dependence of the eccentricity at binding on the mass ratio (see Fig.\,\ref{fig:ebound}). We caution however that our models and setup are significantly different.

The coalescence timescales obtained combining the $N$-body models with a semi-analytical treatment of the late hardening and GW phase range from a few hundred Myr to a few Gyr, in full agreement with previous studies \citep[e.g.][]{preto2011,  khan2011}. We caution that our semi-analytic estimation of the coalescence time-scales ignores rotation of the merger remnant, a key driver of binary hardening. Earlier studies have shown that BHBs corotating with surrounding stellar shroud tends to circularise and those in counter rotation attain very high eccentricities approaching unity \citep{Mirza2017, Varisco2021}. Additionally, it has been observed that BHB hardening rates are higher in rotating environment \citep{holley+15}.

\section{Summary and conclusions}
We have studied the eccentricity of BHBs formed in galactic mergers from early times to the hardening phase, with particular attention to the phase of binary formation. This phase has received little attention in the past due to its intrinsically chaotic nature that makes it difficult to model. Our main results are:
\begin{itemize}
\item Binary formation is an extended phase characterised by strong oscillations in the orbital elements, and a good prescription for the binary formation time is the time when the stellar mass enclosed in the binary orbit is $10\%$ of the BHB total mass.
\item The eccentricity of the BHB, as measured by the pericentre and apocentre of the orbit, tends to decrease during the merger phase due to the action of dynamical friction against the stars. For the case of 1:10 minor mergers, we confirm this behaviour with a simple semi-analytical model following the inspiral of a massive point mass satellite in the fixed gravitational potential of the primary galaxy. For a fixed density profile, larger initial eccentricities lead to faster orbital decay due to the larger efficiency of dynamical friction at pericentre and the larger dissipation experienced deeper into the stellar cusp. Similarly, for a fixed initial eccentricity, the inspiral is slower in steeper density profiles due to the reduced number of stars in the outer regions.
\item Following the dynamical friction phase, the eccentricity remains largely unchanged in the mergers with mass ratio $q=0.25$ and $q=1$, while it increases in the $q=0.1$ mergers. This is in qualitative agreement with semi-analytical models of BHB evolution bases on 3-body scattering experiments.
\item Modelling the later hardening of the BHB due to stellar encounters and GW emission semi-analytically gives total coalescence time-scales of $\sim 300$ Myr for a Milky Way type galaxy and central MBH, and $\sim 100-500$ Myr for a galaxy with a $10^8 \msun$ MBH.
% The later hardening of the BHB due to stellar encounters and GW emissions can be modelled semi-analytically, giving total coalescence time-scales of $\sim 300$ Myr for a Milky Way type galaxy and central MBH, and $\sim 100-500$ Myr for a galaxy with a $10^8 \msun$ MBH.
\item The main parameter determining the eccentricity at binary formation is the initial eccentricity of the merger, with only a minor dependence on the slope of the density profile and the mass ratio.
\end{itemize}

We note that these results hold in the absence of rotation in the merging galaxies. Shorter merger timescales are expected in rotating systems due to more efficient hardening. We will investigate the effects of rotation on the eccentricity of massive black hole binaries at formation in a future work.

\section*{Acknowledgements}
FK and KHB were supported through NASA ATP Grant 80NSSC18K0523.
EB and AS acknowledge support from the European Research Council (ERC) under the European Union's Horizon 2020 research and innovation program ERC-2018-COG
under grant agreement N.~818691 (B~Massive).
MB acknowledges funding from MIUR under the grant PRIN 2017-MB8AEZ.
PB acknowledges support by the Chinese Academy of Sciences (CAS) through the Silk Road Project at NAOC and 
the President’s International Fellowship (PIFI) for Visiting Scientists program of CAS.
The work of PB was also supported by the Volkswagen Foundation under the Trilateral Partnerships grants No.~90411 and~97778 and 
under the special program of the NRF of Ukraine "Leading and Young Scientists Research Support" - "Astrophysical Relativistic Galactic Objects (ARGO): life cycle of active nucleus", No.~2020.02/0346.
PB also acknowledges the support from the Science Committee of the Ministry of Education and Science of the Republic of Kazakhstan (Grant No. AP08856149).

\section*{Data availability}
The data underlying this article will be shared on reasonable request to the corresponding author.



%\bibliographystyle{mnras}

\bibliographystyle{mnras}
\bibliography{biblio}

% Don't change these lines
\bsp	% typesetting comment
\label{lastpage}
\end{document}